%% file: enrichment_distribution.tex
\documentclass[a4paper,fleqn,usenatbib]{mnras}

\usepackage[T1]{fontenc}
\usepackage{ae,aecompl}

\usepackage{graphicx}	
\usepackage{amsmath}	
\usepackage{amssymb}	
\usepackage{xspace}

\numberwithin{equation}{section}

\newcommand{\Rdisk}{$R_{\mathrm{disk}}$\xspace}
\newcommand{\Mdisk}{$M_{\mathrm{disk}}$\xspace}

\newcommand{\al}{$^{26}$Al\xspace}
\newcommand{\fe}{$^{60}$Fe\xspace}

\newcommand{\alz}{Z$_{\mathrm{Al}}$\xspace}
\newcommand{\fez}{Z$_{\mathrm{Fe}}$\xspace}

\newcommand{\alzss}{Z$_{\mathrm{Al,SS}}$\xspace}
\newcommand{\fezss}{Z$_{\mathrm{Fe,SS}}$\xspace}

\newcommand{\alr}{$^{26}$Al/$^{27}$Al\xspace}
\newcommand{\fer}{$^{60}$Fe/$^{56}$Fe\xspace}

\newcommand{\alfe}{$^{26}$Al/$^{60}$Fe\xspace}

\newcommand{\Msun}{$M_{\odot}$\xspace}
\newcommand{\Mstar}{$M_{\star}$\xspace}
\newcommand{\rcl}{r$_{\mathrm{cl}}$\xspace}
\newcommand{\rmhalf}{r$_{\mathrm{M/2}}$\xspace}

\newcommand{\nbody}{$N$-body\xspace}
\newcommand{\nt}{$10^3$\xspace}
\newcommand{\nf}{$10^4$\xspace}
\newcommand{\ec}{$\eta_{\mathrm{cond}}$\xspace}
\newcommand{\eg}{$\eta_{\mathrm{geom}}$\xspace}
\newcommand{\ei}{$\eta_{\mathrm{inj}}$\xspace}
\newcommand{\qui}{{\em quiescent}\xspace}

\title[Isotopic enrichment of forming planetary systems from supernova pollution]{Isotopic enrichment of forming planetary systems from supernova pollution}

\author[T. Lichtenberg et al.]{
Tim Lichtenberg,$^{1,2}$\thanks{Corresponding author, email: tim.lichtenberg@phys.ethz.ch}
Richard J. Parker,$^{3}$
and Michael R. Meyer$^{2}$
\\
$^{1}$Institute of Geophysics, ETH Z{\"u}rich, Sonneggstrasse 5, 8092 Z{\"u}rich, Switzerland\\
$^{2}$Institute for Astronomy, ETH Z{\"u}rich, Wolfgang-Pauli-Strasse 27, 8093 Z{\"u}rich, Switzerland\\
$^{3}$Astrophysics Research Institute, Liverpool John Moores University, 146 Brownlow Hill, Liverpool, L3 5RF, UK
}

\date{Accepted 2016 August 1. Received 2016 July 18; in original form 2016 May 20}

\pubyear{2016}

\begin{document}
\label{firstpage}
\pagerange{\pageref{firstpage}--\pageref{lastpage}}
\maketitle

\begin{abstract}


Heating by short-lived radioisotopes (SLRs) such as \al and \fe fundamentally shaped the thermal history and interior structure of Solar System planetesimals during the early stages of planetary formation. The subsequent thermo-mechanical evolution, such as internal differentiation or rapid volatile degassing, yields important implications for the final structure, composition and evolution of terrestrial planets. SLR-driven heating in the Solar System is sensitive to the absolute abundance and homogeneity of SLRs within the protoplanetary disk present during the condensation of the first solids.
In order to explain the diverse compositions found for extrasolar planets, it is important to understand the distribution of SLRs in active planet formation regions (star clusters) during their first few Myr of evolution.
By constraining the range of possible effects, we show how the imprint of SLRs can be extrapolated to exoplanetary systems and derive statistical predictions for the distribution of $^{26}$Al and $^{60}$Fe based on $N$-body simulations of typical to large clusters ($10^3$-$10^4$ stars) with a range of initial conditions. 
We quantify the pollution of protoplanetary disks by supernova ejecta and show that the likelihood of enrichment levels similar to or higher than the Solar System can vary considerably, depending on the cluster morphology. Furthermore, many enriched systems show an excess in radiogenic heating compared to Solar System levels, which implies that the formation and evolution of planetesimals could vary significantly depending on the birth environment of their host stars.

\end{abstract}



\section{Introduction}
\label{sec:introduction}

The presence of short-lived radioisotopes (SLRs) during the early stages of planetary formation is of central importance in the view of core-accretion planet formation models. In the Solar System, the radioactive decay of \al was the main heat source of the earliest planetesimals and planetary embryos during the first few Myr \citep{1993Sci...259..653G} after the formation of Calcium-Aluminum-rich inclusions (CAI). Their interior thermo-mechanical evolution facilitated differentiation and thus mineralogical, petrographical and structural evolution \citep{2006M&PS...41...95H,2011MaPS...46..903M,lichtenberg16a} and possibly affected the total volatile budget \citep{1999Sci...286.1331Y,2014E&PSL.390..128F}. Subsequent collisional interactions during runaway growth shaped the core-to-mantle ratio and determined the building material of terrestrial planets like Earth \citep{2008RSPTA.366.4205O,2015Icar..247..291B}. With differing initial abundances of SLRs and thus heating rates many of these mechanisms would change and potentially result in drastically different planetary compositions.

SLR tracers in the meteoritic record additionally provide stringent constraints on the birth environment of the nascent Solar System \citep{1976GeoRL...3..109L}. Some of them, like $^{10}$Be, can be explained by solar energetic particle irradiation, but others, most importantly \al and \fe, suggest an external, stellar nucleosynthetic source. To complicate the picture, their concentrations are too elevated to be consistent with galactic background levels \citep{2000SSRv...92..133M}, with the short half-lives (0.7 Myr for \al, 2.6 Myr for \fe) implying a late stage enrichment.

Numerous attempts have been made to link the inferred SLR abundance levels to a specific injection channel including triggered collapse of the presolar cloud core \citep[][]{1977Icar...30..447C}, potentially with former enrichment by the winds of a massive star in a sequential triggering process \citep[][]{2010ApJ...714L..26T,2012A&A...545A...4G,2014E&PSL.392...16Y}, or the direct pollution of the circumstellar disk by supernova ejecta \citep{1977Icar...32..255C,2000ApJ...538L.151C}. The difficulties in measuring excess abundances of \fe, the strongest argument for a direct supernova injection mechanism, so far failed to converge on a preferred order of magnitude \citep{2003ApJ...588L..41T,2010ApJ...720.1215Q,2012EaPSL.359..248T,2016EaPSL.436...71M}. While triggered star formation is a much-debated issue \citep{2015MNRAS.450.1199D}, we consider intra-cluster enrichment mechanisms, like direct disk injection from the winds of nearby massive stars or supernova ejecta, to be inevitable to some extent.

Even though the specific enrichment channel of the Solar System is still under much debate \citep[e.g.,][]{2014prpl.conf..809D,2014ApJ...789...86A,2014MNRAS.437..946P,2015GeCoA.149...88S,2015ApJ...809..103B,2016MNRAS.456.1066P}, attempts are ongoing to extrapolate the predictions by the proposed enrichment channels to extrasolar and even galactic scales \citep{2015A&A...582A..26G}. In the context of the rapidly evolving field of exoplanetary studies \citep{2014prpl.conf..691B} the injection efficiency and hence distribution of SLRs on larger scales link the geodynamical processes in forming planetary systems to its stellar birth environment.

Altough many implications of SLR-dominated heating in planetesimals of sizes greater than $\sim$10 km are now understood \citep[e.g.,][]{2006M&PS...41...95H,2011E&PSL.305....1E,2012AREPS..40..113E,2013AREPS..41..529W,2014M&PS...49.1083G,lichtenberg16a}, the detailed coupling between interior thermo-mechanical-chemical evolution, collisional growth and subsequent effects on the final planet outcome remains elusive \citep{2014E&PSL.390..128F,2015Natur.527..221G,2015Icar..247..291B,2015ApJ...813...72C}. Nonetheless, with the goal of an observationally consistent theory of planetary assembly in mind, it is crucial to understand the distribution of SLRs on larger, like interstellar or galactic, scales.

Thus, in this work, we derive statistical predictions for the distribution of \al and \fe, the two main nucleosynthetic SLR heat sources, from the direct pollution of circumstellar disks in young star clusters. The structure of the paper is as follows. In Sect. \ref{sec:methodology} we describe the \nbody simulations and the post-processing calculations, which take into account the dynamical evolution of the star cluster as well as the injection and mixing of supernova ejecta in the disks, as they evolve over time. We present our results in Sect. \ref{sec:results}, focusing first on the dynamical evolution of the cluster populations, and secondly the predictions for the SLR distribution in them. We discuss the results and limitations of our study in Sect. \ref{sec:discussion} and comment on the implications of our study on volatile degassing, planet population synthesis and the enrichment of the Solar System. We draw conclusions in Sect. \ref{sec:conclusions}.


\section{Methodology}
\label{sec:methodology}

Our method to derive predictions for the SLR abundances in star forming regions is built on a large suite of \nbody simulations of stellar clusters from \nt--\nf stars. This corresponds to the cluster mass required to form stars massive enough to explode as supernovae during the protoplanetary disk phase \citep{2009ApJ...696L..13P,2010ARA&A..48...47A}, assuming a fully sampled initial mass function \citep{2007MNRAS.380.1271P,2010MNRAS.401..275W}. 

We use \nbody simulations with stellar evolution to determine the distances to the supernova(e) of all low-mass stars as a function of the initial conditions of the cluster. We then analyse these simulations using a post-processing routine, which semi-analytically treats details of the pollution mechanism and introduces assumptions about protoplanetary disk lifetimes and isotope mixing. In Sect. \ref{sec:cluster_setup} we present the setup and reasoning of the numerical simulations and explain the post-processing routine in Sect. \ref{sec:enrichment_mechanism}.

\subsection{Star cluster setup}
\label{sec:cluster_setup}


Our model clusters initially have either \nt or \nf stars drawn from the initial mass function of stars (IMF) from  \citet{2013MNRAS.429.1725M}, with a mass range of $M_{\star} = 0.01 - 50 \, M_{\odot}$. This combines the log-normal approximation from \citet{2003PASP..115..763C} with the \citet{1955ApJ...121..161S} power-law slope for stellar masses $> 1 \, \textrm{M}_{\odot}$. The probability density function for the Maschberger IMF has the form
\begin{align}
p(m) \propto \left( \frac{m}{\mu} \right)^{-\alpha} \left( 1 + \left( \frac{m}{\mu}  \right)^{1-\alpha} \right)^{-\beta},
\end{align}
with the average stellar mass $\mu = 0.2 \, \mathrm{M}_{\odot}$, the Salpeter power-law exponent $\alpha = 2.3$ for higher mass stars and $\beta = 1.4$ is used to determine the slope of the low-mass part of the distribution. 

We adopt two different spatial morphologies for the initial conditions of our star clusters; smooth and substructured. The spatial distribution of stars in older clusters is often observed to be smooth and centrally concentrated. These clusters can often be approximated with a \citet{1911MNRAS..71..460P} or \citet{1966AJ.....71...64K} profile.  

Formally, Plummer spheres are infinite in extent, and are usually described in terms of their half-mass radius. We use Plummer spheres with positions and velocities determined by the prescription in \citet{1974A&A....37..183A}, with initial half-mass radii of 0.3 and 0.4 pc.

Stars are observed to form in filaments \citep[e.g.][]{2014prpl.conf...27A}, which usually results in a substructured spatial and kinematic distribution for stars in a given star-forming region \citep{2004MNRAS.348..589C,2009ApJ...696.2086S}. 

We set up star-forming regions with primordial substructure using the fractal distribution in \citet{2004AaA...413..929G}. Note that we are not claiming that star-forming regions are fractals \citep[although they may be,][]{2001AJ....121.1507E}, but rather that fractals are the most convenient way of setting up substructure because the amount of substructure is set by just one number, the fractal dimension $D$.

For a detailed description of the fractal set-up, we refer the interested reader to \citet{2010MNRAS.407.1098A} and \citet{2014MNRAS.438..620P}, but we briefly summarise it here. The fractal is built by creating a cube containing `parents', which spawn a number of `children' depending on the desired fractal dimension. The amount of substructure is then set by the number of children that are allowed to mature. The lower the fractal dimension, the fewer children are allowed to mature and the cube has more substructure. Fractal dimensions in the range $D = 1.6$ (highly substructured) to $D = 3.0$ (uniform distribution) are allowed. Finally, outlying particles are removed so that the cube from which the fractal was created becomes a sphere; however, the distribution is only truly spherical if $D = 3.0$. We adopt $D = 1.6$ throughout this paper, and the fractals have initial radii of 1\, and 3\,pc. 

To determine the velocity structure of the cloud, children inherit their parent's velocity plus a random component that decreases with each generation of the fractal. The children of the first generation are given random velocities from a Gaussian of mean zero. Each new generation inherits their parent's velocity plus an extra random component that becomes smaller with each generation. This results in a velocity structure in which nearby stars have similar velocities, but distant stars can have very different velocities. The velocity of every star is scaled to obtain the desired virial ratio of the star-forming region.

The virial ratio,  $\alpha_{\mathrm{vir}} = T/|\Omega|$, where $T$ and $|\Omega|$ are the total kinetic and total potential energy of all stars in the cluster, respectively. \citet{2014MNRAS.437..946P} found that the amount of enrichment is insensitive to the initial virial ratio in the range $\alpha_{\mathrm{vir}} = 0.3$ (bound and initially collapsing) to $\alpha_{\mathrm{vir}} = 0.7$ (unbound with some initial expansion). Therefore, all clusters were initialised in virial equilibrium, with $\alpha_{\mathrm{vir}} = 0.5$.

Our model clusters are evolved using the {\sc kira} integrator within the {\sc starlab} package \citep[see, e.g.,][]{1999A&A...348..117P,2001MNRAS.321..199P} and we implement stellar evolution using the {\sc seba} code \citep[][]{1996A&A...309..179P,2012ascl.soft01003P}, also in {\sc starlab}. The clusters are evolved for 10\,Myr, which encompasses the majority of protoplanetary disk lifetimes, stars $\gtrsim$\,19\,M$_\odot$ will explode as supernovae and the majority of dynamical interactions that could disrupt disks occur within this timeframe.

\begin{table}
\centering
\caption{Initial settings for all types of $N$-body simulations, which we follow for 10 Myr in total. All simulations are initialised using $D = 1.6$ and $\alpha_{\mathrm{vir}} = 0.5$. From left to right, the columns are the number of stars,  $N_{\star}$, the initial morphology, and either the initial radius of the fractal $R_{\mathrm{cl}}$, or the initial half-mass radius of the Plummer sphere $r_{1/2}$. To account for statistical variations we perform 30 simulation runs for each type.}
\label{tab:sim_parameters}
\begin{tabular}{ccc}
\hline
$N_{\star}$ & Morphology & $R_{\mathrm{cl}}$ or $r_{1/2}$ [pc]  \\
\hline
10$^3$ & Fractal & 1.0 \\
10$^3$ & Plummer & 0.3 \\
10$^3$ & Plummer & 0.4 \\
10$^4$ & Fractal & 1.0 \\
10$^4$ & Fractal & 3.0 \\
10$^4$ & Plummer & 0.3 \\
10$^4$ & Plummer & 0.4 \\
\hline
\end{tabular}
\end{table}


To account for the stochasticity inherent in star cluster evolution \citep{2012MNRAS.424..272P,2014MNRAS.437..946P}, we ran 30 versions of each set of initial conditions, identical apart from the random number seed used to generate the masses, positions and velocities. A summary of the initial settings can be found in Tab. \ref{tab:sim_parameters}. Finally, each simulation was run for 10 Myr, the higher end tail of protoplanetary disk lifetimes and thus the relevant timescale for supernova pollution of circumstellar disks \citep[see, e.g.,][]{2001ApJ...553L.153H,2003ARA&A..41...57L,2009AIPC.1158....3M,2013MNRAS.434..806B,2014ApJ...796..127C}.

\subsection{Enrichment mechanism}
\label{sec:enrichment_mechanism}

\begin{figure}
    \centering
	\includegraphics[width=\columnwidth]{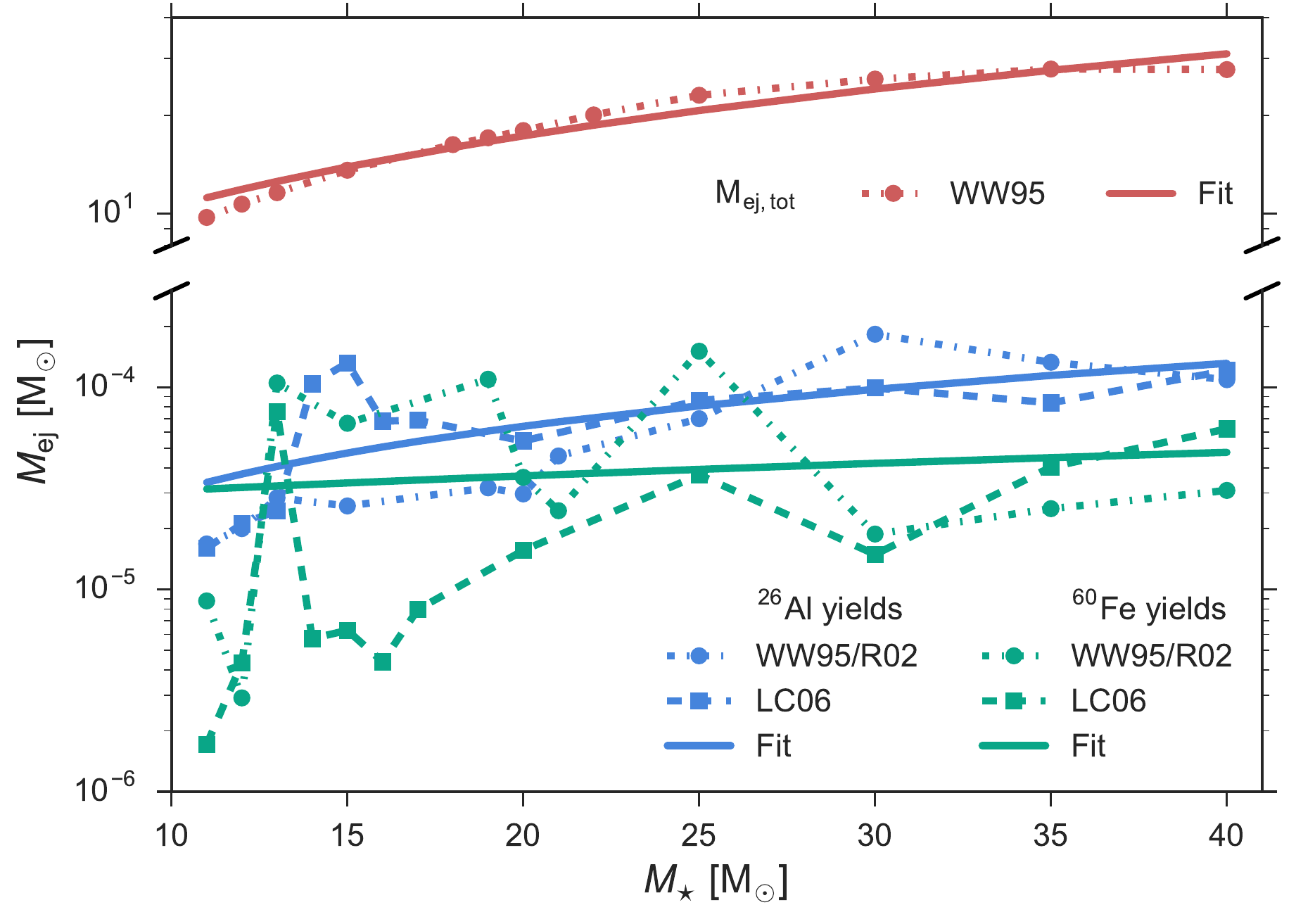}
    \caption{Isotopic yields of \al and \fe per supernova progenitor mass with data from \citet[][]{1995ApJS..101..181W,2002ApJ...576..323R} (denoted as WW95/R02) and \citet[][]{2006ApJ...647..483L} (LC02). To achieve a yield function we fit the data linearly. The regressions scale as: $M_{\mathrm{ej,tot}} \propto 0.68 M_{\star}$, $M_{\mathrm{ej,Al}} \propto 3.4 \times 10^{-6} M_{\star}$ and $M_{\mathrm{ej,Fe}} \propto 5.6 \times 10^{-7} M_{\star}$.}
    \label{fig:progenitor_yields}
\end{figure}

It is common for young stars to form close together in a cluster and therefore supernova progenitor stars are found in the vicinity of less massive stars \citep{2003ARA&A..41...57L}. After several Myr of evolution the progenitor(s) explode as core-collapse supernovae, ejecting their SLR-enriched outer shells into the local environment. Parts of the ejecta material are injected into circumstellar disks via cross-section capture, mixed into the disk and subsequently incorporated into forming solid bodies. In the following we describe each physical process from the SLR-ejection of supernova progenitor stars down to the mixing in the disk and present our post-process implementation in order to derive statistical predictions for the distribution of SLRs in stellar clusters with \nt -- \nf stars.

First, we use the \nbody simulations to determine the locations of stars when a massive star goes supernova and the relative positions of low-mass stars in the cluster with respect to the supernova. By fitting the supernova yield predictions from \citet[][]{1995ApJS..101..181W,2002ApJ...576..323R,2006ApJ...647..483L} we assign each supernova event to a specific absolute mass of SLR material, as shown in Figure \ref{fig:progenitor_yields}.

Using the formulation of \citet{2007ApJ...662.1268O,2010ApJ...711..597O}, the transport and injection physics of supernova yields to a disk can be broken down into 3 parameters: the condensation efficiency of metals in the ejector shock front \ec, the cross-section factor of a protoplanetary disk with the ejector front \eg and the injection efficiency into the disk \ei. \citet{2010ApJ...711..597O} discussed the observational constraints of \ec and arrived on a possible parameter range of \ec = 0.01--1. Since then, however, evidence of efficient solid condensation in ejector fronts has arisen  \citep{2011Sci...333.1258M} and it seems indeed possible that \ec is close to $\sim$1. To take a conservative approach and to account for potential variations between different supernova events we chose \ec = 0.5 throughout our analysis. The geometrical cross-section of a random disk with supernova ejecta can be easily calculated \citep{2014ApJ...789...86A} via
\begin{align}
\eta_{\mathrm{geom}} = \frac{\pi r_{\mathrm{disk}}^2}{4 \pi d^2}\cos{\theta}, \label{eq:eta_geom}
\end{align}
with $r_{\mathrm{disk}}$ the radius of the circumstellar disk (see Sect. \ref{sec:timing}), $d$ the distance between donor and acceptor star and $\theta = 60^{\circ}$ the average disk alignment with respect to the ejector for a random distribution. For this purpose we neglected the time interval $dt$ between the supernova outburst and the time of arrival, as average travel times in such a scenario are of the order of $dt \approx 100$ yr \citep{2010ApJ...711..597O}, i.e., negligible in comparison with the evolutionary timescale of the disks, their half-lives and the cluster dynamical timescale.

The injection efficiency \ei is determined by the size distribution of condensed grains in the supernova ejecta, as large grains can be easier injected. For the case of a grain size distribution in accordance with presolar grains of supernova origin, like meteorite grain sizes in the Solar System  \citep{1994GeCoA..58..459A}, the injection efficiency into the disk structures is $\approx
$ 0.9 \citep{2010ApJ...711..597O}. However, we assume that the average dust grain size is smaller, $\sim 0.1 \mu$m, consistent with the size distribution of interstellar grains \citep{1977ApJ...217..425M}, in which case \ei$\approx 0.7$ \citep{2009GeCoA..73.4946O}. In general, whenever we had to decide between an approach which would boost final enrichment abundances versus one which is expected to lower them, we chose the latter, which we see as a conservative way of not overestimating the effects of direct supernova pollution.

When the SLRs are injected into the disk they must be mixed into the disk material on a relatively short time scale as is implicitly assumed in when using \al as a Solar nebula chronometer in cosmochemic dating methods. From a theoretical perspective, mixing in the disk can occur either via large scale gravitational instabilities or turbulent diffusion on local scales. The first is extremely effective while the disk is massive \citep[e.g.,][]{2007ApJ...660.1707B}. The latter demands a very detailed understanding of the actual disk physics and crucially depends on still largely unconstrained disk properties, but can also be efficient when the vertical structure of the disk is considered due to large-scale radial mixing \citep{2006Icar..181..178C,2007Sci...318..613C}. A detailed review of the mixing processes within the disk is beyond the scope of this paper, for further details see \citet{2009GeCoA..73.4946O}. For the Solar System, there is evidence of large-scale \al homogeneity \citep{2009Sci...325..985V}, even though the discussion is still ongoing.  When the SLRs are mixed into the disk material before planetesimal agglomeration, they will be incorporated into the growing bodies.

\subsection{Timing and disk dynamics}
\label{sec:timing}

\begin{figure}
    \centering
    \includegraphics[width=\columnwidth]{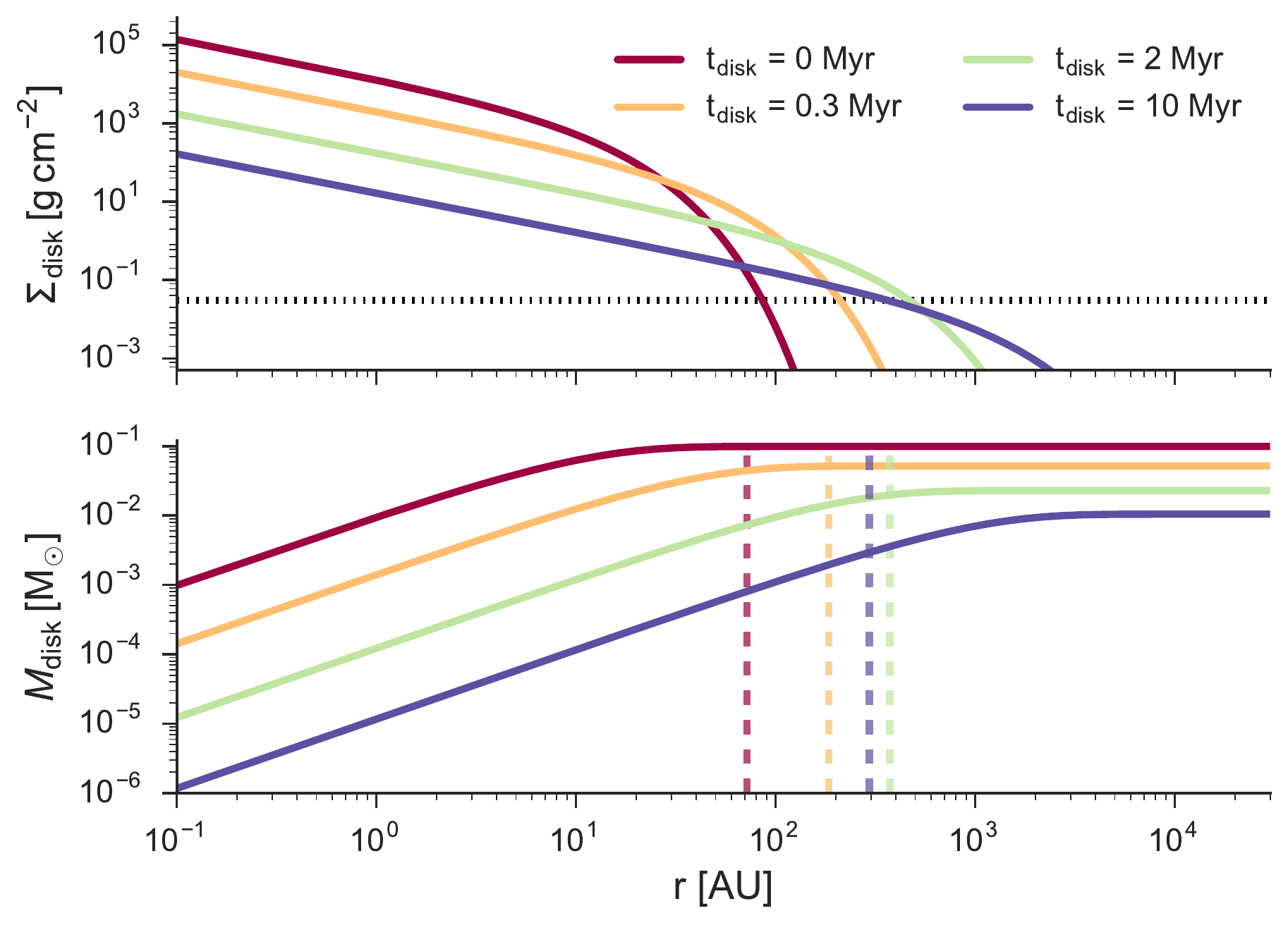}
    \caption{Time dependent disk model \citep{2009apsf.book.....H} to calculate the truncation radius \Rdisk and total mass \Mdisk for a Solar analogue with $M_{\star} = 1 \, M_{\odot}$. {\bf Upper panel:} Surface density for different times. The horizontal dotted line indicates the cut off density from the density criterion (compare Fig. \ref{fig:disk_truncation}). {\bf Lower panel:} Disk mass for different times with the truncation radii (again from the density criterion) for different times indicated with vertical dashed lines.}
    \label{fig:disk_model}
\end{figure}

\begin{figure*}
    \centering
    \includegraphics[width=2.0\columnwidth]{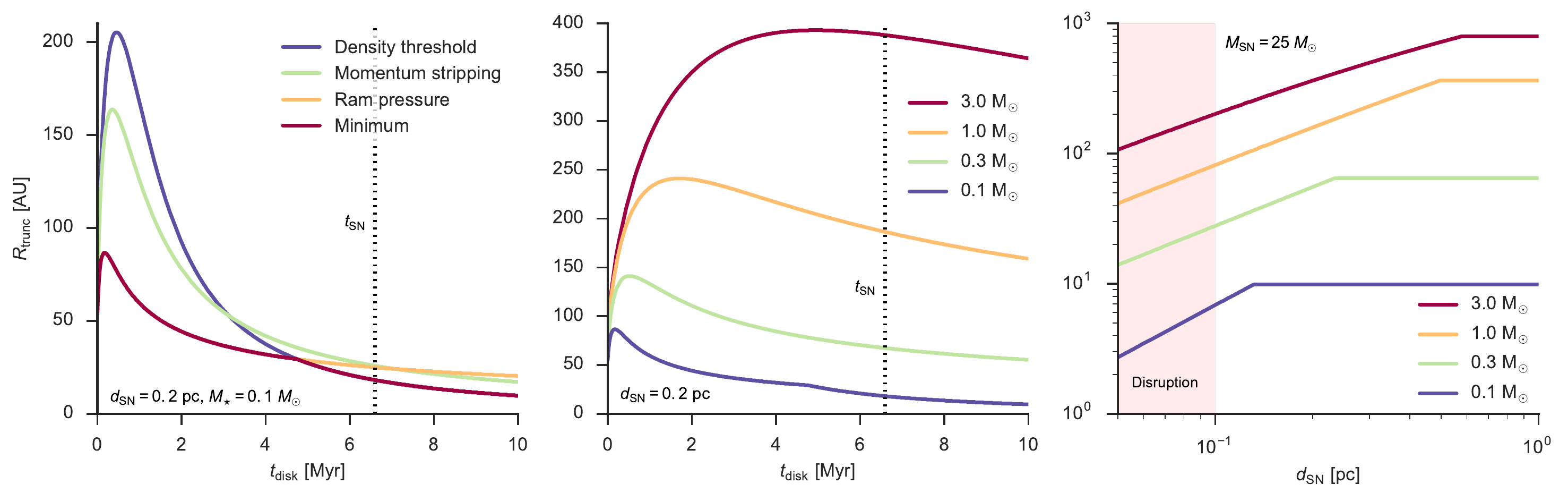}
    \caption{Disk truncation radius \Rdisk, calculated from the minimum value of the estimates for momentum stripping, ram pressure and density threshold (compare Fig. \ref{fig:disk_model}). {\bf Left:} Truncation radius for fixed distance from a supernova and mass of the disk host star. During the early evolution of the disk the truncation is determined by the amount of ram pressure, whereas from $\sim$5 Myr onwards momentum stripping and the density threshold become the limiting factors. {\bf Middle:} Disk truncation radius for various host star masses. The kink in the slope for a host star of $0.1 M_{\odot}$ results from the transition from ram pressure dominated truncation to density threshold truncation as can be seen in the left image. {\bf Right:} Truncation radii $t = 6.6$ Myr, the explosion time for a supernova with progenitor mass $M_{\mathrm{SN}} = 25 M_{\odot}$. The truncation radii increase with increasing distance to the supernova, as momentum and ram stripping become less rigorous, and are capped from the density threshold criterion.}
    \label{fig:disk_truncation}
\end{figure*}

\begin{table*}
\centering
\caption{List of physical parameters in the numerical $N$-body model and analytical post-processing routine. Parameters without reference are further described in Sect. \ref{sec:methodology}.}
\label{tab:parameters}
\begin{tabular}{lllll}
\hline
Parameter & Symbol & Value & Unit & Reference \\
\hline
\hline
Cluster fractal dimension & $D$ & 1.6 & & \citet{2004AaA...413..929G} \\ 
Cluster virial ratio & $\alpha_{\mathrm{vir}}$ & 0.5 & & \\
Cluster fractal radius & $R_{\mathrm{F}}$ & 1.0/3.0 & pc & \\
Cluster half-mass radius & $R_{\mathrm{1/2}}$ & 0.3/0.4 & pc & \\
Simulation time & $t_{\mathrm{sim}}$ & 10 & Myr & \\
Number of stars & $N_{\star}$ & $10^3$/$10^4$ & & \\
\hline
Supernova explosion energy & $E_{\mathrm{ej}}$ & $1.2 \times 10^{51}$ & ergs & \citet{2007ApJ...662.1268O,2002ApJ...576..323R}   \\
Dust condensation in supernova ejecta & \ec & 0.5 &  & \citet{2010ApJ...711..597O,2011Sci...333.1258M}  \\
Ejector-disk cross-section capture & \eg & Eq. \ref{eq:eta_geom} &  & \citet{2009GeCoA..73.4946O,2010ApJ...711..597O}  \\
Injection efficiency & \ei & 0.7 &  & \citet{2009GeCoA..73.4946O,2010ApJ...711..597O}  \\
\hline 
Initial disk radius & $R_1$ & 10 & AU & \citet{2009apsf.book.....H} \\
Initial disk mass & $M_d(0)$ & 0.1 $M_{\star}$ & M$_{\odot}$ & \citet{2009apsf.book.....H} \\
Disk alignment & $\theta$ & 60 & $^{\circ}$ & \\
Disk viscosity & $\alpha$ & $10^{-2}$ & & \citet{2009apsf.book.....H,2015ApJ...813...99F} \\
Disk normalisation temperature & $T_{100}$ & 10 & K & \citet{2009apsf.book.....H} \\
Disk cut-off density & $\Sigma_{\mathrm{out}}$ & 0.03 & g cm$^{-2}$ & \citet{2007ApJ...662.1268O,2010ApJ...711..597O} \\
Disk dust-to-gas ratio &  & 0.01 &  &   \\
\hline 
Al concentration in CI chondrites & $f_{\mathrm{Al,CI}}$ & $8.5 \times 10^{3}$ & $\mu$g g$^{-1}$ & \citet{2003ApJ...591.1220L}, Tab. 3 \\ 
\alr Solar System\footnote{Upper-limit value.} & \alzss & $5.85 \times 10^{-5}$ & & \citet{2006ApJ...646L.159T} \\ 
\al decay energy & $E_{\mathrm{Al}}$ & $3.12$ & MeV & \citet{2009Icar..204..658C}  \\
\al half-life & $t_{\mathrm{1/2,Al}}$ & $7.17 \times 10^{5}$ & yr & \citet{2009Icar..204..658C}  \\
\al radiogenic heating at CAI & $Q_{\mathrm{Al,SS}}$ & $1.7 \times 10^{-7} $ & W/kg &  \\
Fe concentration in CI chondrites & $f_{\mathrm{Fe,CI}}$ & $182.8 \times 10^{3}$ & $\mu$g g$^{-1}$ & \citet{2003ApJ...591.1220L}, Tab. 3 \\ 
\fer Solar System, low & \fezss & $1.15 \times 10^{-8}$ & & \citet{2012EaPSL.359..248T} \\ 
\fer Solar System, high & & $\sim 10^{-6}$ & & \citet{2016EaPSL.436...71M} \\ 
\fe decay energy & $E_{\mathrm{Fe}}$ & $2.712$ & MeV & \citet{2009Icar..204..658C}  \\
\fe half-life & $t_{\mathrm{1/2,Al}}$ & $2.60 \times 10^{6}$ & yr & \citet{2015PhRvL.114d1101W}  \\
\fe radiogenic heating at CAI & $Q_{\mathrm{Fe,SS}}$ & $8.325 \times 10^{-11} $ & W/kg & based on low \fe value \\
Combined radiogenic heating at CAI & $Q_{\mathrm{r,SS}}$ & $1.7 \times 10^{-7} $ & W/kg & \citet{2011MaPS...46..903M} \\
\hline
\end{tabular}
\end{table*}

We assumed that every star forms with a disk of initially \Mdisk = 0.1 \Mstar, which is expected to be the treshold mass just stable to large-scale gravitational instabilities at the end of the infall phase \citep{2000prpl.conf..559N,2011ARA&A..49...67W}. The initial number of potential planet forming disks was therefore $N_{\mathrm{disk}} = N_{\star}$, depending on the cluster size. As we were interested in the consequences for systems which finally developed fully fledged planetary systems, we subtracted the stars which might have been subject to violent transformation due to either disruption by close-by supernova \citep[$d \leq 0.1$ pc,][]{2010ARA&A..48...47A}, truncation by close encounters with $d \leq 500$ AU or photoevaporation from a nearby O star \citep[$d \leq 0.3$ pc for $dt \geq 1$ Myr,][]{2001MNRAS.325..449S,2004ApJ...611..360A,2014prpl.conf..475A}. Again, we have chosen these values in a conservative fashion, as not to overpredict SLR abundances. See Sect. \ref{sec:limitations} for a discussion of these issues.

The number of stars/disks not violated by these events was subsequently scaled by a declining distribution, to account for the decreasing number of observed disks with cluster age \citep{2001ApJ...553L.153H,2009AIPC.1158....3M}. To do so, at the time of each supernova event, the number of disks was decreased via the exponential scaling law
\begin{align}
N_{\mathrm{disk}} = N_{\star} e^{-t/\tau_{\mathrm{disk}}},
\end{align}
with mean lifetime $\tau_{\mathrm{disk}} = 5.0$ Myr, corresponding to a disk half-life time of $\tau_{\mathrm{1/2}} = \tau_{\mathrm{disk}} \ln{(2)} \approx 3.47$ Myr. Given that newest observational estimates are consistent with up to $\tau_{\mathrm{1/2}} \approx 6.0$ Myr \citep{2013MNRAS.434..806B,2014ApJ...796..127C}, this is again a conservative approach.

To achieve a realistic estimate of the mixing ratio of the SLRs with the disk material, we inferred a time-dependent disk model for the evolution of the surface density \citep{2009apsf.book.....H} with a flared structure according to
\begin{align}
& T(R) \sim 10 \left( \frac{100 \, \mathrm{AU}}{R} \right)^{1/2} \mathrm{K}, \\
\label{eq:hartmannA}
& \Sigma \sim 1.4 \cdot 10^3 \frac{e^{-R/(R_1 t_d)}}{(R/R_1)t_d^{3/2}} \left( \frac{M_d(0)}{0.1 \, \mathrm{M}_{\odot}} \right) \left( \frac{R_1}{10 \, \mathrm{AU}} \right)^{-2} \, \mathrm{g\,cm}^{-2}, \\
& t_d = 1 + \frac{t}{t_s}, \\
& t_s \sim 8 \cdot 10^4 \left( \frac{R_1}{10 \, \mathrm{AU}} \right) \left( \frac{\alpha}{10^{-2}} \right)^{-1} \left( \frac{M_{\star}}{0.5 \, M_{\odot}} \right)^{1/2} \left( \frac{T_{100}}{10 \, \mathrm{K}} \right) \, M_{\odot} \, \mathrm{yr}^{-1},
\label{eq:hartmannZ}
\end{align}
with $R$ the distance from the central star, the `initial' scaling disk radius $R_1 = 10$ AU\footnote{This is \textit{not} a total initial disk radius, compare Fig. \ref{fig:disk_model}.}, time normalisation $t_d$, scaling time $t_s$, initial disk mass $M_d(0) = 0.1 \, M_{\star}$, disk viscosity parameter $\alpha = 10^{-2}$ and normalisation temperature $T_{100} = 10$ K. For a detailed description of this model we refer to \citet{2009apsf.book.....H}. Even though newest theoretical and observational estimates of the disk viscosity trend towards lower values, our choice of $\alpha = 10^{-2}$ reflects a compromise between the seemingly low values in observed disks \citep{2014prpl.conf..475A,2015ApJ...813...99F} and long disk lifetimes for lower $\alpha$ values in this model.

As the disks became subject to interaction with supernova blast waves and could potentially be truncated from the momentum coupling, we inferred an estimate of the disk truncation radii during the intersection with the ejecta. In a simple analytical estimate \citet{2000ApJ...538L.151C} stated that two types of mass loss by the interaction with the ejecta must be taken into account. First, matter may be stripped from a disk due to the ram pressure of the ejector flow, if the timescale of the ejecta interaction is longer than the dynamical timescale of the disk. Then, the disk is stripped if the ram pressure exceeds the gravitational force per unit area
\begin{align}
P_{\mathrm{grav}} \approx G M_{\star} \Sigma_{\mathrm{disk}} / r_{\mathrm{d}},
\end{align}
with the gravitational constant $G$, the surface density of the disk $\Sigma_{\mathrm{disk}}$ at distance $r_{\mathrm{d}}$. For the case of a constant density supernova with ejection energy $E_{\mathrm{ej}}$, \citet{2000ApJ...538L.151C} found the maximum ram pressure that can be exerted by the ejecta to be 
 \begin{align}
P_{\mathrm{ram}} = 5 E_{\mathrm{ej}} / (2 \pi d_{\mathrm{SN}}^3),
\end{align}
with the distance to the supernova source $d_{\mathrm{SN}}$. Second, in the case of a rapid interaction of the ejecta with the disk, the momentum in the ejecta flow $p_{\mathrm{m,ej}}$ can cause disk material to reach escape velocity
\begin{align}
v_{\mathrm{esc}} = (2 G M_{\star} / r_{\mathrm{d}})^{1/2}.
\end{align}
Thus, the momentum stripping criterion becomes
\begin{align}
& p_{\mathrm{m,ej}} > p_{\mathrm{m,disk}},\\
& M_{\mathrm{ej}} v_{\mathrm{ej}} / (4 \pi d_{\mathrm{SN}}^2) > \Sigma_{\mathrm{disk}} v_{\mathrm{esc}},
\label{eq:mom_strip}
\end{align}
with $v_{\mathrm{ej}} = (10 E_{\mathrm{ej}}/ 3 M_{\mathrm{ej}})^{1/2}$. However, \citet{2007ApJ...662.1268O} found that these criteria are too restrictive because disks in interaction with the ejecta become surrounded by high-pressure shocked gas that cushions the disk and deflects gas-phase ejecta around it. The associated bow shock deviates the gas and lowers the effectiveness of momentum stripping and ram pressure by orders of magnitudes. In their model, which is inspired by potential Solar System initial conditions, $d_{\mathrm{SN}} = 0.1$pc, $M_{\star} = 25$ AU, and the gas is stripped beyond the radius with surface density $\Sigma \sim 0.03$ g cm$^{2}$ at $\sim$33 AU. 

Therefore, we employed a semi-analytical, iterative approach to calculate the disk truncation radius for each injection event. We began with a time-dependent disk profile when the supernova explodes and the ejector hits the disk. An exemplary disk model for the profiles inferred from Eq. \ref{eq:hartmannA} for a Solar analogue is shown in Figure \ref{fig:disk_model}, together with the surface density threshold criterion from \citet{2007ApJ...662.1268O}. 

For this setup we iteratively computed from inside-out the ratio of ejecta and disk momentum $p_{\mathrm{m,ej}} / p_{\mathrm{m,disk}}$ from  Eq. \ref{eq:mom_strip}, compared this with the momentum stripping ratio for the parameter combination of \citet{2007ApJ...662.1268O}. For the ram pressure we followed a similar approach, and iteratively computed from inside-out the ratio of ejecta ram pressure to the gravitational force per unit area, and compared this with corresponding ratio calculated from the values in \citep{2007ApJ...662.1268O}. 
Finally, we chose the minimum truncation radius inferred from the momentum stripping, ram pressure and density cut-off. The disk truncation radius and thus the final mass of the disk were therefore dependent on the specific supernova as well as the disk structure. A demonstration for a subset of parameters is shown in Fig. \ref{fig:disk_truncation}. The supernova explosion energy $E_{\mathrm{ej}} = 1.2 \times 10^{51}$ ergs, as was found for SN 1987A \citep{2002ApJ...576..323R}. In principle, this value can vary with progenitor mass, but \citet{2007ApJ...662.1268O} found that the differences in mass loss rates from different values for $E_{\mathrm{ej}}$ are insignificant for the energy range in \citet{1995ApJS..101..181W,2002ApJ...576..323R}.

Finally, to calculate mixing ratios and derive heating values, we assumed Solar-like compositions of all disks, with a dust-to-gas ratio of 0.01, listed in Tab. \ref{tab:parameters}. We calculated the ratios for \al and \fe, derived the ratios \alr and \fer and from this calculated an initial heating ratio for planetesimals under the assumption of instantaneous homogeneous mixing and planetesimal formation with chondrititic compositions \citep{2003ApJ...591.1220L}. The heating ratio from radioactive decay of SLRs can be computed from \citep{2011MaPS...46..903M}
\begin{align}
Q_{\mathrm{r}}(t) = f_{\mathrm{Al,CI}} \left[ \frac{^{26}\mathrm{Al}}{^{27}\mathrm{Al}} \right] \frac{E_{\mathrm{Al}}}{\tau_{\mathrm{Al}}} e^{-t/\tau_{\mathrm{Al}}} + f_{\mathrm{Fe,CI}} \left[ \frac{^{60}\mathrm{Fe}}{^{56}\mathrm{Fe}} \right] \frac{E_{\mathrm{Fe}}}{\tau_{\mathrm{Fe}}} e^{-t/\tau_{\mathrm{Fe}}},
\end{align}
with the chondritic abundance of Al or Fe $f_{\mathrm{Al/Fe,CI}}$, the decay energy of \al and \fe, $E_{\mathrm{Al/Fe}}$ and mean lifetime of both radioisotopes, $\tau_{\mathrm{Al/Fe}}$, respectively. Note that we have chosen an upper-limit estimate of the initial Solar System \alr ratio, as given in \citet{2006ApJ...646L.159T}.

If there were more than one supernova event in the simulation, we calculated the remaining SLR material from the former supernova event from the isotopic half-life time and added up the remnant material and the new contribution from the current supernova. For the statistics derived in in the results, we accepted the SLR abundance event for each individual with the {\it maximum} heating rate from the combined effect of \al and \fe. Therefore, our results reflect the maximum value of SLR abundance each individual system received over the lifetime of the simulation. See Sect. \ref{sec:discussion} for a discussion of the implications of this.



\section{Results}
\label{sec:results}

\begin{figure*}
    \centering
    \includegraphics[width=0.99\textwidth]{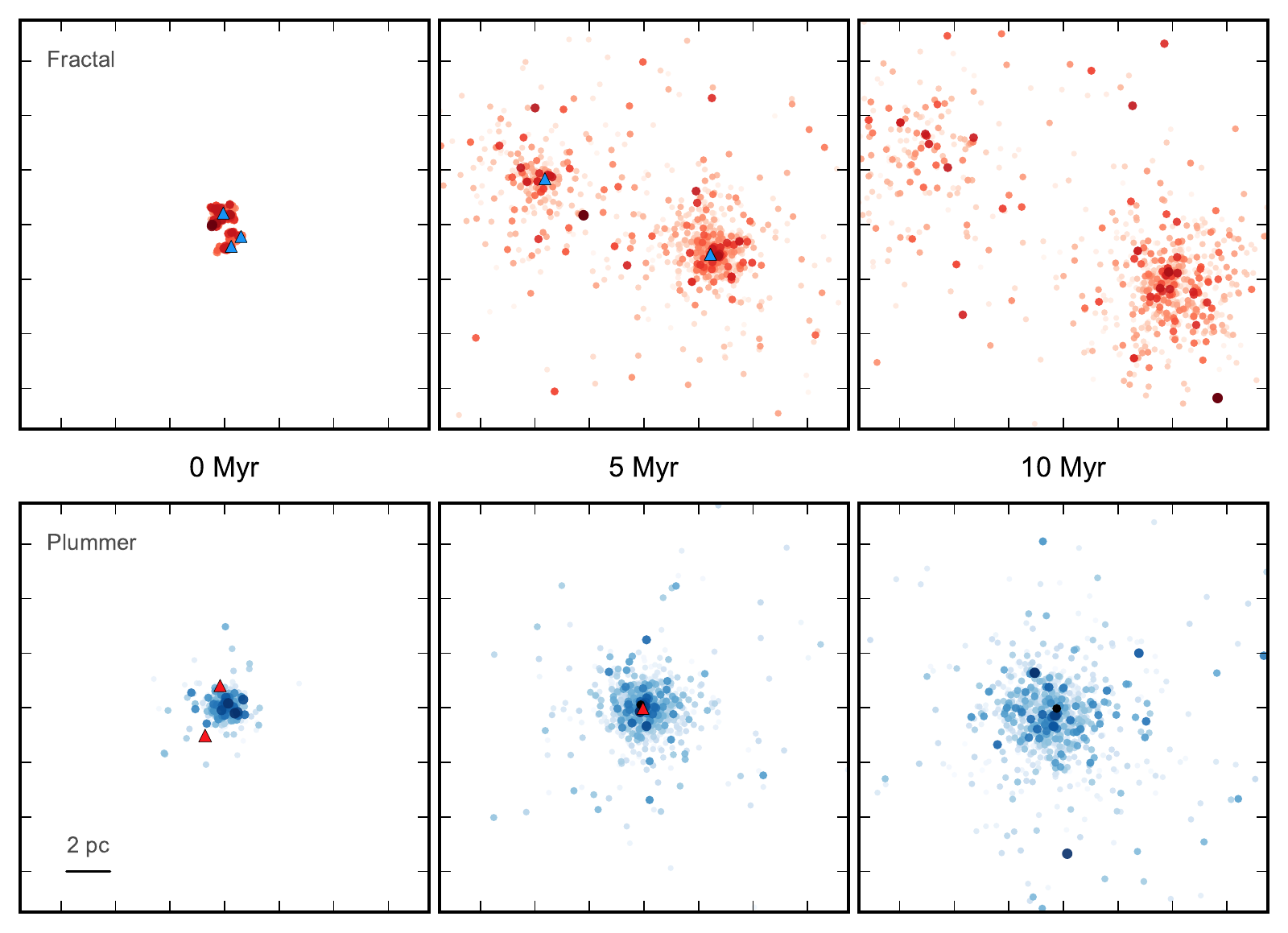}
    \caption{Two-dimensional projection of the dynamical evolution of example stellar clusters with 10$^3$ stars. Stars $\gtrsim 19$ \Msun, shown as triangles, explode as supernovae within $t < t_{\mathrm{sim}}$ and are therefore gone at $t = 10$ Myr. Smaller stars are shown as circles, with color brightness scaling with stellar mass. {\bf Upper panel:} Fractal simulation with \rcl = 1 pc. The initially highly substructured cluster formed two separate subclusters after a few Myr of  evolution, each containing supernova progenitor stars. {\bf Lower panel:} Plummer simulation with $r_{1/2} = 0.3$ pc. This simulation retained its spherically symmetric morphology and developed a clear mass segregation signature.}
    \label{fig:scatter_e3}
\end{figure*}

\begin{figure*}
    \centering
    \includegraphics[width=0.99\textwidth]{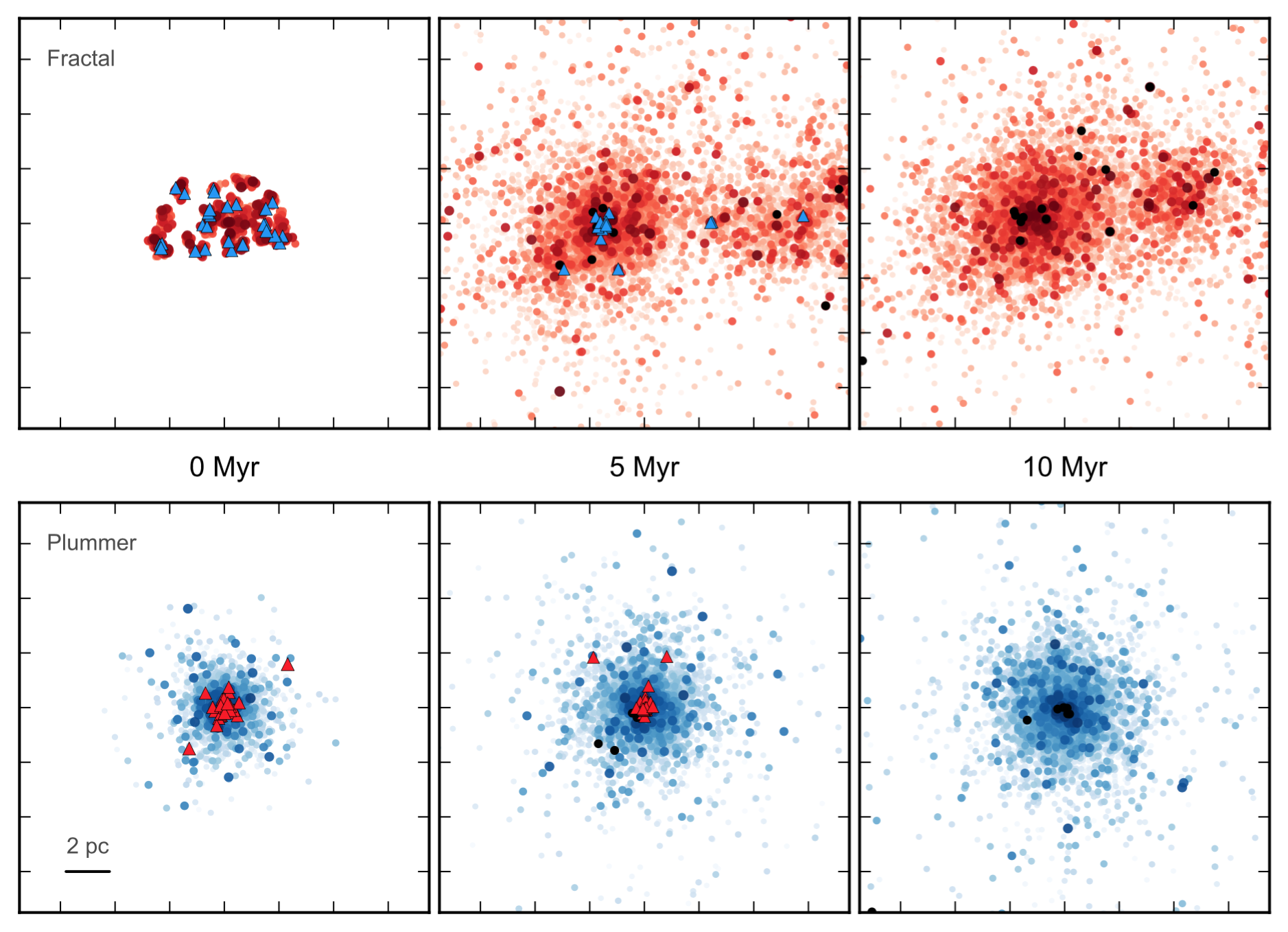}
    \caption{Two-dimensional projection of the dynamical evolution of example stellar clusters with 10$^4$ stars. Stars $\gtrsim 19$ \Msun, shown as triangles, explode as supernovae within $t < t_{\mathrm{sim}}$ and are therefore gone at $t = 10$ Myr. Smaller stars are shown as circles, with color scaling with stellar mass. {\bf Upper panel:} Fractal simulation with \rcl = 3 pc. The initially highly substructured cluster showed a complicated morphology with supernova progenitor stars in very dense and less dense regions. {\bf Lower panel:} Plummer simulation with \rmhalf = 0.4 pc. Although dynamical mass segregation usually leads to the most massive stars residing in the central region, some massive stars exploded as supernovae in the outskirts of the cluster.}
    \label{fig:scatter_e4}
\end{figure*}

\begin{figure}
	\centering
	\includegraphics[width=\columnwidth]{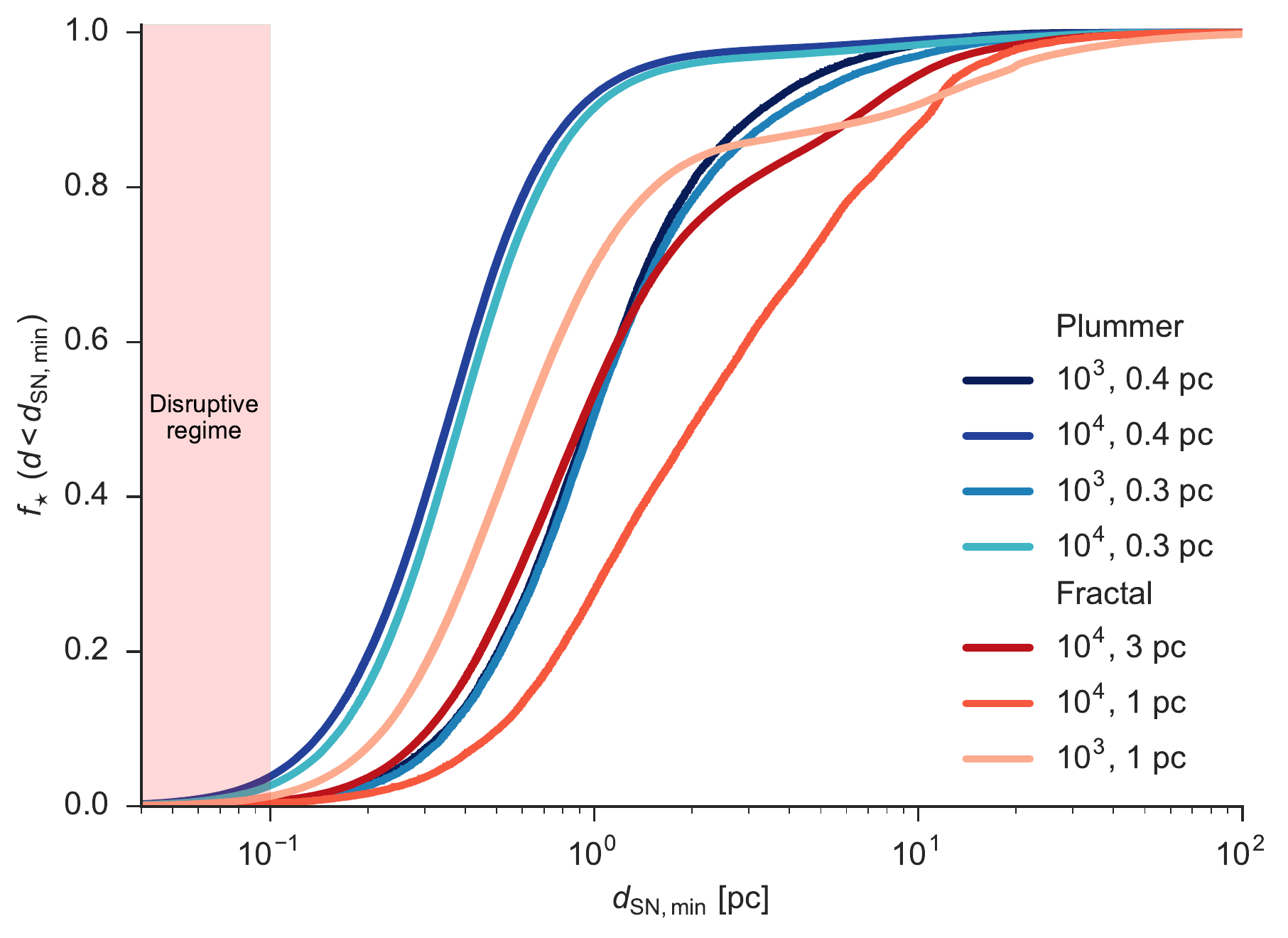}
    \caption{Cumulative minimal distances to a supernova event during $t_{\mathrm{sim}}$ time for each star in simulations with at least one event. Stars within the red zone were within 0.1\,pc of a supernova and could potentially have their disks destroyed, so we subtracted them from the enrichment analysis. 
    On average, stars within $10^4$ star-clusters lay closer to supernova progenitors than to stars in $10^3$ star-clusters. Furthermore, fractal morphologies increased the average distance in comparison with Plummer type clusters.}
    \label{fig:distances_type}
\end{figure}

\begin{table*}
	\centering
	\caption{Averaged results of the \nbody simulations, with the number of of supernova events during runtime $N_{\mathrm{SN}}$, the number of disks disrupted by supernova $N_{\mathrm{disrupt}}$, the number of disks perturbed by close stellar encounters $N_{\mathrm{perturb}}$, the number of disks potentially violently altered by photoevaporation of near-by O stars $N_{\mathrm{evap}}$ and the number of disks subject to at least one these effects $N_{\mathrm{destroy}}$. $N_{\mathrm{destroy}}$ could be lower than the sum of the former three values, as a single disk could be subject to more than one effect.}
	\label{tab:cluster_results}
	\begin{tabular}{lll||rrrrr}
	\hline
	$N_{\star}$ & Morphology & $R_{\mathrm{cl}}$ or $r_{1/2}$ [pc] & $N_{\mathrm{SN}}$ & $N_{\mathrm{disrupt}}$ [\%] & $N_{\mathrm{perturb}}$ [\%] & $N_{\mathrm{evap}}$ [\%] & $N_{\mathrm{destroy}}$ [\%] \\
	\hline
	\input{./tabs/tab_cluster_results.txt}
	\hline
	\end{tabular}
\end{table*}

In this section we present the results of our study. First, in Sect. \ref{sec:cluster_dynamics}, we describe details of the \nbody simulations and discuss the dynamical aspects of the enrichment mechanism. Second, in Sect. \ref{sec:enrichment_distribution} we present the results from the post-processing of the simulations, deriving predictions for the distribution of \al and \fe and resulting radiogenic heating rates.

\subsection{Star cluster dynamics}
\label{sec:cluster_dynamics}

Figures \ref{fig:scatter_e3} and \ref{fig:scatter_e4} illustrate the dynamical evolution of $10^3$ and $10^4$ simulations with different number of stars and cluster morphologies for snapshots at times $t=$ 0, 5 and 10 Myr. The supernova progenitor stars were randomly distributed throughout the simulations in the initial conditions. After $t=$ 5 Myr most clusters underwent some degree of dynamical mass segregation, such that the massive stars resided at the cluster center or at the center of a sub-cluster structure (as in the fractal simulation in Figure \ref{fig:scatter_e3}). At $t=$ 10 Myr all of the stars $> 19$ $M_{\odot}$ have exploded as supernovae, losing most of their mass and becoming remnant objects, i.e., a black hole or neutron star. For the enrichment distribution, as presented in Sect. \ref{sec:enrichment_distribution}, it is very important to note that the progenitor stars were not perfectly segregated into the middle of the cluster. Instead, there were often stars in the outskirts of the cluster, thus enriching stars far away from the cluster center.
The fractal simulations often did not form radially symmetric density distributions. Following the erasure of some of the initial substructure they dynamically evolve into association-like complexes with dense subgroups, where the massive stars usually reside approximately at the center of these subgroups \citep[compare][]{2014MNRAS.438..620P}.

Figure \ref{fig:distances_type} illustrates the effect of the cluster morphology on the spatial configuration of stars with respect to a supernova progenitor star. In this plot we show the shortest distance $d_{\mathrm{min}}$ of each star in the simulations to a supernova event for all timesteps. This supernova event was very likely to determine the enrichment outcome for the specific star as the enrichment cross-section scales with $\sim d^{-2}$ (see Eq. \ref{eq:eta_geom}). We highlight the `disruptive' zone ($d_{\rm SN} < 0.1$pc) for circumstellar disk evaporation by the shaded red region. If stars were within this distance of the supernova event they were classified as disrupted by the enormous energy injection of the ejecta shock front \citep{2000ApJ...538L.151C}. Stars in Plummer-type clusters on average show smaller  $d_{\mathrm{min}}$ than their fractal counterparts. The reason for this is that a Plummer-sphere relaxes via two-body dynamics only, whereas the fractals initially relax via violent relaxation \citep{1967MNRAS.136..101L}, which leads to more drastic expansion \citep{2012MNRAS.427..637P}, and therefore higher $d_{\mathrm{min}}$ values. In total, $\sim 5-40$ \% of stars lie in the expected zone for high pollution efficiencies $0.1 < d_{\rm SN} < 0.3$ pc \citep{2010ARA&A..48...47A}.

However, the distance to the closest supernova was not enough to derive the enrichment distribution. First of all, as described in Sect.~\ref{sec:timing}, the stars may have been subject to other violent interactions, like perturbations by close encounter with other stars or by intense mass-loss due to external photoevaporation by nearby O-type stars. We quantify our analysis of these effects in Tab. \ref{tab:cluster_results}. In the Plummer-sphere simulations, low-mass stars were rarely subject to perturbing close encounters. In the fractals $\sim 3-4$ \% of all stars were subject to these interactions.

Photoevaporation by the aggressive radiation fields of O-type stars in the clusters, however, turned out to be an influential mechanism on the survival of disks using our formulation. The fraction of disks subject to potentially evaporative radiation varied from $\sim$17--79\% and strongly depended on the initial cluster morphology. Here, stars in fractal clusters tended to be less influenced on average, as they spent less time in photoevaporation zones. Plummer geometries turned out to be more hostile than fractals, because up to 4 out of 5 stars were potentially affected by photoevaporation. It is worth emphasising that stochastic differences in the dynamical evolution result in a large spread in these values between clusters \citep[see also][]{2012MNRAS.424..272P,2014MNRAS.437..946P,2014MNRAS.438..620P}.

The very last column indicates the fraction of disks subject to at least one of the former effects and therefore gives the reverse of the number of disks we considered as {\em quiescent}, meaning they did not suffer from close encounters, ejecta disruption of photoevaporation effects as defined in Sect. \ref{sec:timing}. These discs could be enriched, if they were not already dispersed `naturally' (by random drawing from the declining distribution, see Sect. \ref{sec:timing}) by the time of a supernova event. These were the disks for which we derived the enrichment distribution, as described in the next section.

\subsection{Enrichment distribution}
\label{sec:enrichment_distribution}


\begin{figure}
	\centering
    \includegraphics[width=\columnwidth]{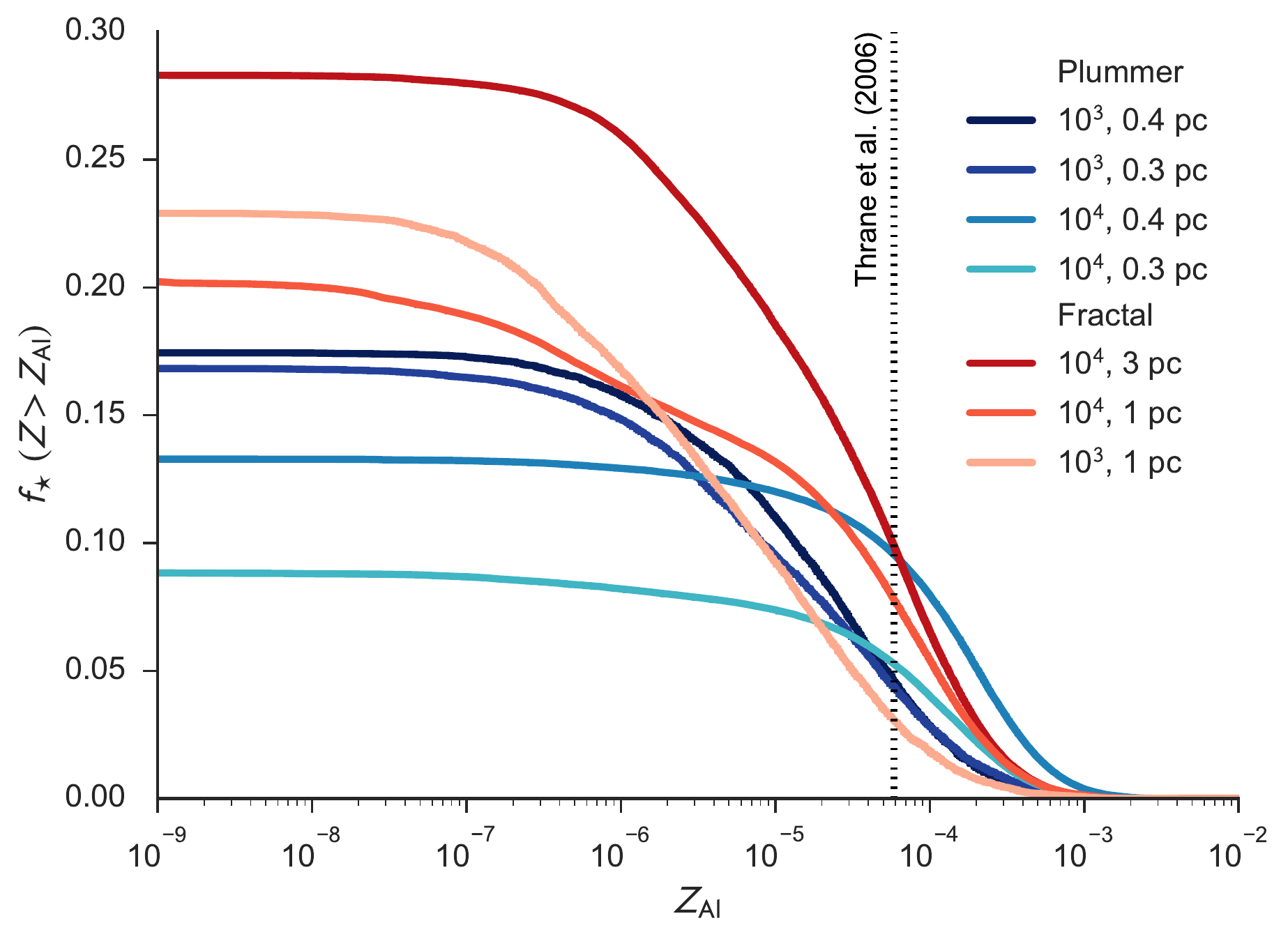}
    \includegraphics[width=\columnwidth]{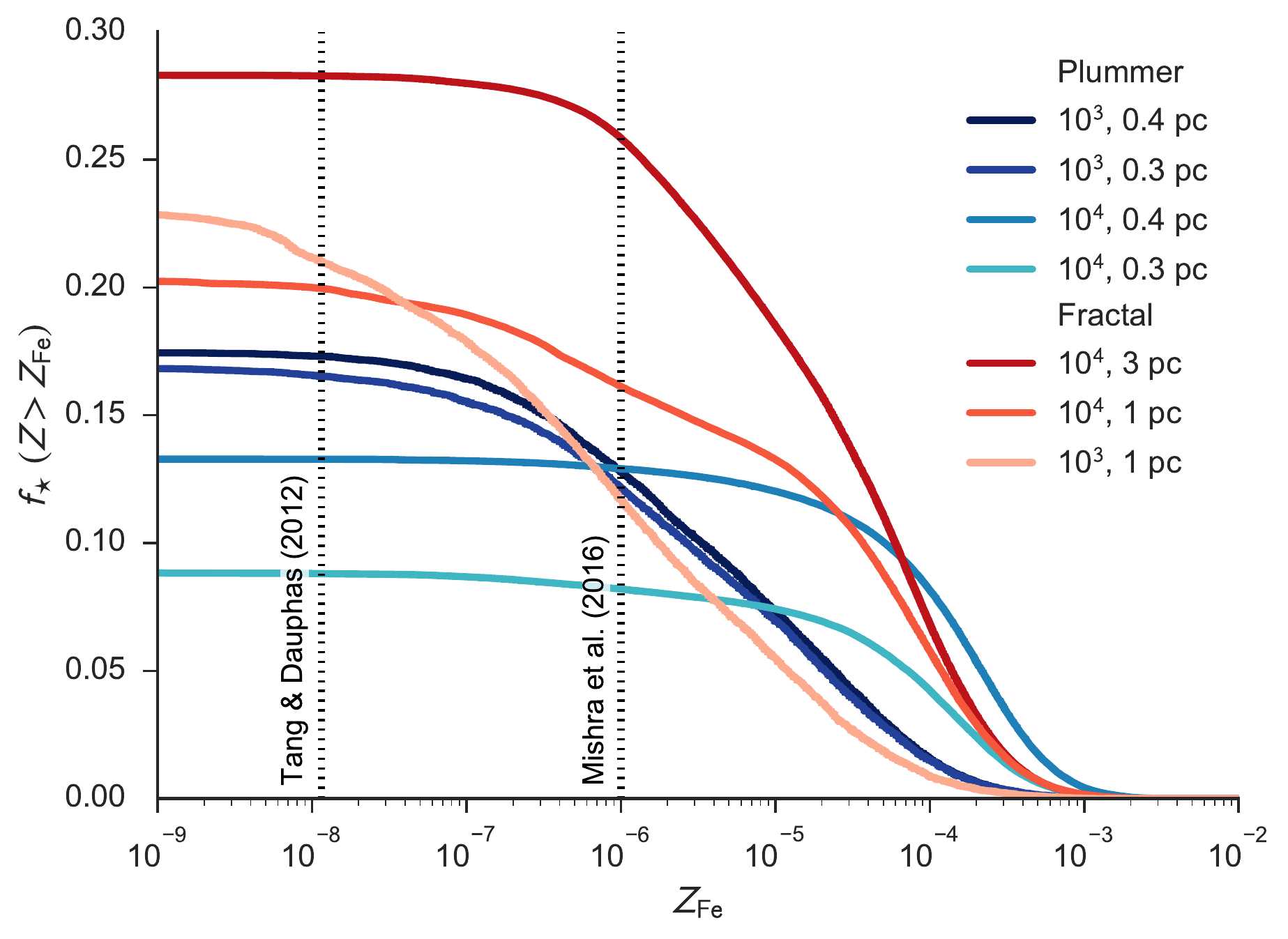}
    \caption{Inverse cumulative \al and \fe yield distributions for stars in clusters, derived in terms of isotopic mixing ratios \alz (\alr) and \fez (\fer) within disks of initially Solar compositions. The vertical dotted line shows the Solar System values. $\sim$3--10 \% of disks have a similar or higher \al abundance to the Solar System. For \fe the abundance was typically higher, with $\sim$8--28 \% of disks bearing \fez $\gtrsim$ \fezss.}
    \label{fig:hist_enr1}
\end{figure}

\begin{table*}
    \centering
    \caption{Enrichment statistics. All values are scaled to the total number of stars at the {\em beginning} of the simulations. $\mathrm{f}_{\mathrm{enr}}$ gives the fraction of disks without violent disruption, perturbation or photoevaporation events (compare Tab. \ref{tab:cluster_results}) {\em and} did survive until the first supernova event within the specific simulation. The next two columns give the average value of <\alz> and <\fez> and <$\mathrm{Q}_{\mathrm{r}}$> the resulting average heat inside primitive planetesimals. $\mathrm{f}_{\mathrm{Q > SS}}$ gives the computed fraction of disks with an average heat higher than the Solar System.}
    \label{tab:enrichment_stats}
    \begin{tabular}{lll|rrrrr}
    \hline
    $N_{\star}$ & Morphology & $R_{\mathrm{cl}}$ or $r_{1/2}$ [pc] & $\mathrm{f}_{\mathrm{enr}}$ [\%] & <\alz> $[10^{-5}]$ & <\fez> $[10^{-5}]$ & <$\mathrm{Q}_{\mathrm{r}}$> [$Q_{\mathrm{r,SS}}$] & $\mathrm{f}_{\mathrm{Q > SS}}$ [\%] \\
    \hline
	\input{./tabs/tab_enrichment_stats.txt}
    \hline
    \end{tabular}

\end{table*}

\begin{figure}
	\centering
    \includegraphics[width=\columnwidth]{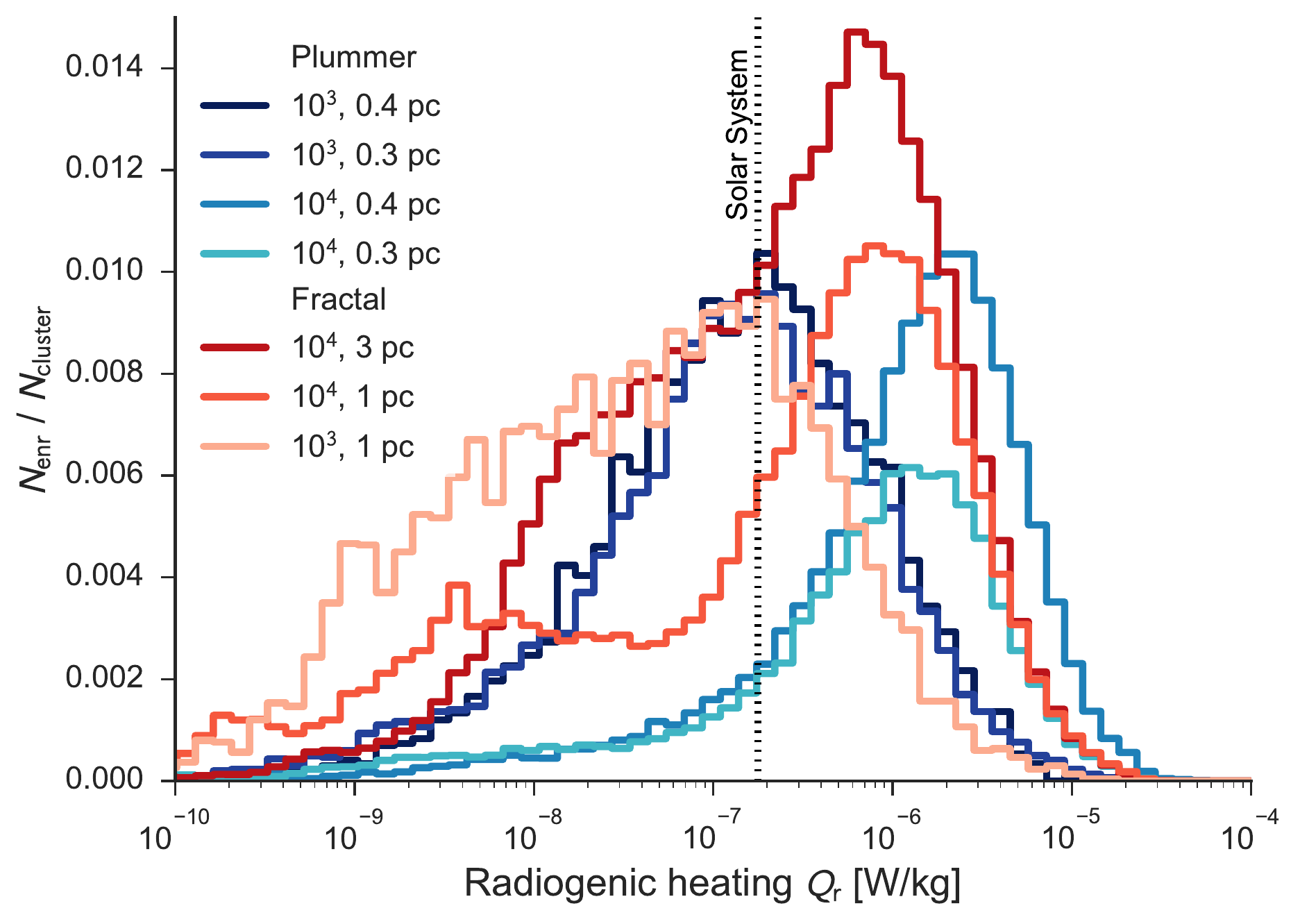}
    \caption{Maximum radiogenic heating in planetesimals in enriched systems directly after injection of supernovae ejecta. The vertical dotted line indicates the Solar System value. If a system was enriched via supernova pollution it was likely that planetesimals in there experienced stronger heating than in the Solar System.}
    \label{fig:hist_enr2}
\end{figure}

In this section we analyse the results for the enrichment distribution for the \qui disks from the former section. 

Figure \ref{fig:hist_enr1} shows the enrichment distribution for \al and \fe in terms of the isotopic mixing ratios \alz (\alr) and \fez (\fer), with the Solar System values indicated for comparison. These plots demonstrate various findings. First of all, the \al abundance of disks with \alz $\gtrsim$ \alzss varied from $\sim$3--10 \% as a function of simulation type. \fe abundances, however, showed variations between $\sim$8--28 \% with \fez $\gtrsim$ \fezss for the two end-member simulation types. This was a consequence of our supernova ejecta model, which assumed isotropic expansion and constant \alfe ratio in the ejecta, which is higher than the Solar System value (see Sect. \ref{sec:ss} for a Solar System-focused discussion of this issue). In general, Plummer morphologies show a lower fraction of enriched disks than the fractal geometries. This is mostly because in Plummer simulations the average stellar velocity relative to the cluster center is slower and therefore stars reside longer in the inner regions of the cluster. The fraction of evaporated disks due to photoevaporation is higher and thus fewer disks survive to be enriched. 
Additionally, clusters with $N = 10^3$ stars show higher abundances than $N = 10^4$ clusters, since in these clusters significantly less O-type stars were present. Although this decreases the enrichment levels, the positive effect of less photoevaporation within the simulations outweighs the negative of the lower number of supernova progenitor stars. Finally, fractal geometries showed a shallower slope toward higher abundance. This can be explained with the higher diversity in evolved clusters at the time when the supernova progenitor stars went supernova. Comparing the morphologies in Figures \ref{fig:scatter_e3} and \ref{fig:scatter_e4} shows that in plummer simulations the stellar density decreased with increasing distance to the cluster center, with the supernova progenitors mostly mass segregated in the center. Fractal geometries could deviate significantly from this, as does the distribution of supernova progenitor stars. As already discussed above, the volume in time subject to photoevaporation is enhanced in comparison with fractal simulations and if mass segregation were perfect, enrichment timing would become a crucial issue. This was one of the central critique points of the supernova pollution model by \citet{2007ApJ...663L..33W} and is attenuated by the results of our fractal simulations, where the issue of a star being in the `right' zone is strongly alleviated.

Table \ref{tab:enrichment_stats} summarizes our quantitative results for the enrichment distributions, with $\mathrm{f}_{\mathrm{enr}}$ the fraction of enriched disks surviving disruption, perturbations, evaporation and dispersion until at least on supernova event has occurred, the average values for \alz and \fez and corresponding heating value in chondritic material <$\mathrm{Q}_{\mathrm{r}}$>. $\mathrm{f}_{\mathrm{Q > SS}}$ indicates the fraction of disks with a higher heating value than the Solar System's initial value (compare Table \ref{tab:cluster_results}). The trends discussed for Figure \ref{fig:distances_type} can be confirmed from the average values. 

The value of $\mathrm{Q}_{\mathrm{r}}$, derived from the combined heating, bridges the gap from the enrichment to the physics of forming planets. Figure \ref{fig:hist_enr2} shows the distribution of initial heating from SLRs among different systems at the time of injection for perfectly effective mixing (no time delay between injection and homogeneous distribution in the system). Two features can be extracted from the histograms. First, $10^3$ clusters show median heating rates around the Solar System value. This means, that if a system was enriched, it was most likely to bear similar heating as Solar System planetesimals at CAI. Second, $10^4$ simulations showed an excess in heating, their peaks are shifted to higher enrichment and thus heating levels. This dichotomy was grounded in the number of supernova events in a simulation. If a system was enriched in a $10^4$ simulation, then it was likely to experience multiple enrichment events and from more massive stars, whose ejecta carried more SLRs with them.


\section{Discussion}
\label{sec:discussion}

In the following we discuss and interpret the outcome of our analysis. In Sect. \ref{sec:implications_distribution} we focus on global aspects of the derived distribution dichotomy, with potential effects on the planet formation process and the global planet population in Sect. \ref{sec:implications_planets}, while in Sect. \ref{sec:ss} we compare our results with earlier work with regards to the enrichment mechanism of the Solar System. In Sect. \ref{sec:limitations} we discuss the limitations and uncertainties of our study.

\subsection{Enrichment distribution}
\label{sec:implications_distribution}

As indicated in Sect. \ref{sec:introduction}, the implications of the abundance levels of SLRs are numerous, as they dominated the internal heat budget of planetesimals from $\sim$10--1000 km in the first few Myr after CAI formation in the Solar System. Our results in Tab. \ref{tab:enrichment_stats} and Fig. \ref{fig:hist_enr2} hint on an SLR distribution dichotomy in forming planetary systems. Enriched systems are very likely to bear high concentrations of SLRs and the derived heating value for the early System is by no means the upper limit in those systems. Therefore, if a system is enriched, the heating rate from SLRs $Q_{\mathrm{r}}$ can be several orders of magnitude higher than in the Solar System. Such extreme heating values are unknown from Solar System studies and the full implications remain elusive. As well as the enriched systems, we find a large fraction of systems with zero enrichment. This is partially based in our simplistic acceptance criteria for disrupted or photoevaporated systems (see Sect. \ref{sec:limitations}), but in general agrees with statistical distributions from other studies \citep{2009ApJ...696.1854G,2015A&A...582A..26G}.

Given the frequency distribution of star clusters as a function of mass, we expect equal contributions at logarithmic mass intervals to the total mass in stars in the Milky Way galaxy \citep[e.g.,][]{2003ARA&A..41...57L}. In other words, star forming events with richness 10-100 M$_{\odot}$ contribute as much as events with 10$^4$--10$^5$ M$_{\odot}$. We expect these smallest events ($\sim$20 \% of star forming events) to suffer no SLR enrichment at all. Events 100-1000 M${\odot}$ might contain some stars that experience enrichment. And star clusters 10$^4$--10$^5$ M${\odot}$ ($\sim$20 \% of Population I star forming events) will certainly produce some stars with even greater  SLR enrichment. However we expect the dichotomy we observe here to persist when averaged over all star forming events in the Milky Way: most stars will suffer no enrichment, but the minority that do, will often suffer levels of enrichment much higher than that inferred for our Solar System.


\subsection{Implications for planet formation and population synthesis}
\label{sec:implications_planets}

A multitude of implications for varying levels of SLRs was already envisioned in the literature. For instance, non-negligible SLR abundances can provide an energy source for ionizing ambient disk material \citep{2013ApJ...777...28C}. If the flux from cosmic rays is suppressed by stellar winds, SLRs can in fact be the dominant contribution to the ionization rate present and thus be crucial for the angular momentum transfer via the magnetorotational instability \citep{1991ApJ...376..214B}. Therefore, disk evolution could be fundamentally different in the two end-member states of our derived SLR distribution dichotomy, which would affect the internal dynamics and the planet formation process as a whole. 

Once accreted, the temperatures in the earliest planetesimals, bearing noteworthy SLR abundances, rise and interally stored volatile species become subject to melting processes. If, however, early Solar System objects transformed substantially due to SLR heating \citep[see][and references therein]{2015ApJ...804....9C}, then planetary systems with much higher SLR concentration would be altered fundamentally, potentially losing a high fraction of internally stored volatile species, such as water in the form of hydrated silicates, by degassing and other mechanisms. Comparable to \citet{2009ApJ...696.1854G}, however due to fundamentally different mechanisms, we find Solar System levels of SLR enrichment were typical for enriched systems. The chemistry and thermo-mechanical evolution of small accreted planetary objects further alters hydration reactions and the formation of serpentinites and thus the synthesis processes of potential seeds for life, such as organic compounds like primitive amino acids \citep{2011Icar..213..273A,2014ApJ...783..140C,2015ApJ...809....6C,lichtenberg16a}.

In addition to the initial abundance of SLRs, the formation time and orbital separation relative to molecular ice lines \citep[e.g.,][]{2011ApJ...743L..16O} likely determine the primordial composition and heat generation within planetesimals. In the Solar System, the initial conditions determined thermal metamorphism, aqueous activity and abundance of volatiles in these bodies \citep{2015NatCo...6E7444D}. Thus, the volatile components of finally assembled planets in the distribution of planetary systems may vary significantly. These species, like water and carbon dioxide, have a disproportionate influence on the processes such as planetary differentiation and habitability \citep{2007Icar..191..337S,2010ApJ...708.1326F,2014A&A...561A..41A}. In particular, the initial water content crucially alters the solidification of planetary mantles during the magma ocean phase and degassing pathways, which determines the atmospheric composition and ocean depth to a great degree \citep{1985JGR....90..545A,2011Ap&SS.332..359E}. Theoretical models show that excess water concentrations could result in extremely volatile rich system architectures \citep{2015ApJ...804....9C,2015ApJ...807....9M,2016A&A...589A..15S} and would likely alter the conditions for habitability \citep[e.g.,][]{2012ApJ...756..178A,2014ApJ...781...27C,2015ApJ...801...40S,2016Icar..277..215N}. In conclusion, we envisage that our findings indicate a fundamental difference between volatile rich (low initial SLR content) and volatile poor (high initial SLR content) planetary systems with crucial implications for the planetary populations in these systems. 

\subsection{Solar System enrichment}
\label{sec:ss}

Our findings offer new insights into the SLR enrichment channel of the Solar System and can rule out and/or support certain arguments used in the literature for and against specific enrichment channels.

The \fe abundance is still a much-debated issue and laboratory experiments diverge by more than two orders of magnitude from $\sim$$10^{-8}$--$10^{-6}$ \citep{2012EaPSL.359..248T,2014E&PSL.398...90M,2016EaPSL.436...71M}. The abundance (or simply, the existence) of \fe is, however, of fundamental importance for the enrichment of the Solar System via direct disk pollution (the model featured in this paper) and triggered star formation \citep[as in][]{2012ApJ...745...22G,2015ApJ...809..103B} as the \alfe ratio in supernova models differs greatly from the \alfe ratio in the Solar System. In the formulation used in this paper, we assumed that all supernova yields were transported outwards via isotropic and homogeneous ejecta. This is unlikely, as revealed by observations \citep[e.g.,][]{2009ApJ...691..875L,2012ApJ...746..130H} and potential inhomogeneities and anisotropies (for example, via clumps) could have played a fundamental role in enriching young Solar Systems \citep{2010ApJ...711..597O,2012ApJ...756..102P}. Additionally, recent theoretical models of nucleosynthesis in massive stars still suffer from uncertainties in critical nuclear physics \citep{2007PhR...442..269W}. Therefore, current estimates for the production of SLRs in supernovae and Wolf-Rayet winds must be treated with caution, as the divergence in the models is large enough to account for most of the deviation of Solar System values from the predicted values. Because of these uncertainties, the enrichment by one \citep{1977Icar...30..447C} or more \citep{2007Sci...316.1178B} supernovae by any channel cannot be ruled out.

Chronometric dating of U-corrected Pb-Pb absolute ages of chondrules \citep{2012Sci...338..651C} and paleomagnetic measurements of angrites and Semarkona chondrules \citep{2015LPI....46.2516W} suggest a lifetime of the Solar protoplanetary disk of $\sim$ 4 Myr. During the earliest stages ($t \leq 0.3$ Myr) the \alr ratio was likely heterogeneously distributed \citep{2009GeCoA..73.4963K,2012M&PS...47.1948K}, but may have rapidly approached the so-called canonical value of \alr $\approx 5 \times 10^{-5}$ due to efficient mixing processes, which is often used as an argument for self-enriched molecular cloud models \citep[e.g.,][]{2013ApJ...769L...8V,2016ApJ...826...22K}. 

However, it remains controversial whether such models are consistent with the observed absence of significant age spreads in young star forming regions \citep{2011A&A...534A..83R,2011MNRAS.418.1948J,2012A&A...547A..35C}. In general, massive star formation in young star-forming regions is found to be rapid \citep{2000ApJ...530..277E}, both in simulations \citep{2003MNRAS.343..413B,2012MNRAS.424..377D} and observations \citep{2014prpl.conf..219S}. This underlines the importance for considering intra-cluster enrichment processes.

As we have shown in this study, the likelihood of enriching planetary systems on the level of the Solar System is very common among enriched systems. Additionally, our model results differ from those of \citet{2007ApJ...663L..33W} and \citet{2008ApJ...680..781G}, which conclude that supernova polluted systems were a rare event, even when corrected for photoevaporation and co-evolution of low and high mass stars. Moreover, the enrichment levels for \al differ by up to one order of magnitude from those in \citet{2007ApJ...663L..33W}. For this, we identified two main causes. First, in the model by \citet{2007ApJ...663L..33W} massive stars were assumed to be entirely mass segregated in the cluster center. Therefore, the enrichment of planetary systems was limited to a very narrow zone around the cluster center, such that timing became a crucial issue. Even though timing played a role in our models as well, the limitations were much less severe, as massive stars could be found in the cluster outskirts as well, due to dynamical evolution and thus the possible volume/zone of enrichment was greatly enhanced. The major effect could be seen in the deviations between fractal and Plummer geometries in our simulations.  Second, \citet{2007ApJ...663L..33W} used a much smaller disk lifetime, barely in accordance with recent estimates \citep{2014ApJ...796..127C}.

In summary, we conclude that the enrichment channel of the Solar System is anything but clarified and needs further investigation. Especially, hints on potential heterogeneities or late-stage injections in \al and \fe levels in meteorites \citep[like FUN\footnote{\underline{F}ractionation and \underline{U}nidentified \underline{N}uclear isotope properties \citep{1977GeoRL...4..299W}.} CAIs, which exhibit non-radiogenic isotope abundance anomalies and contained little or no $^{26}$Al at the time of their formation; e.g.,][]{2008ApJ...680L.141T} could open up new ways to derive the enrichment history of the Solar System \citep{2007ApJ...655..678Q,2011ApJ...733L..31M,2016EaPSL.436...71M} and need to be synchronized with astrophysical injection mechanism channels, taking into consideration further aspects, like the direct injection of winds from massive stars into protoplanetary disks.

\subsection{Limitations}
\label{sec:limitations}

In this section, we discuss potential limitations of our study. We divide this discussion into two parts: first, we discuss limitations with regards to the simulations and the choices regarding the {\em stellar} parameters; second, we focus on the disk properties and our assumptions regarding the {\em planetary} growth process.


To begin with, we were not able to account for potential age spreads in star formation and cannot investigate triggered star formation scenarios. However, we note that measurements of stellar ages so far are consistent with age spreads up to the order of $\lesssim$ 2 Myr \citep{2011A&A...534A..83R,2011MNRAS.418.1948J,2012A&A...547A..35C}. These limit the reach of triggered star formation \citep{2012ApJ...756..102P} and thus triggered enrichment in general \citep{2015MNRAS.450.1199D,2016MNRAS.456.1066P} but in turn could even enhance the likelihood of supernova pollution, when massive stars can be formed earlier than low-mass stars.

Our simulations did not contain any primordial binary stars, whereas the initial binary fraction in star formation could be high \citep[e.g.,][]{2012MNRAS.421.2025K,2015ApJ...799..155D}. In general, binary stars should be subject to supernova pollution as well. This, however, would demand a much more complicated treatment of the disk dynamics, hence we neglected it in this study.

As discussed in Sect. \ref{sec:ss}, the supernova ejecta were assumed to expand isotropically and homogenously. This limited our ability to predict the outcome of {\em specific} planetary system. Thus, we did not derive predictions for single, isolated systems. Instead, we focused on statistical predictions of a large ensemble of stars. As the total mass output for supernova models and the \alfe ratio in ejecta are in relative agreement between theoretical models \citep{2002ApJ...576..323R,2000ApJS..129..625L}, we argue that the averaging process corrected for the uncertainties in predictions for single systems and thus our averaged predictions in Tab. \ref{tab:enrichment_stats} were not affected. Furthermore, we assumed a dust condensation efficiency of 0.5, which can be subject to changes. However, as discussed in Sect. \ref{sec:enrichment_mechanism}, there is recent evidence for very high condensation efficiencies \citep{2011Sci...333.1258M} and hence we believe this estimate to be reasonable. 

As found by many authors \citep{1998ApJ...499..758J,1999ApJ...515..669S,2001MNRAS.325..449S,2004ApJ...611..360A} photoevaporation from O-type stars in stellar clusters can severely alter the structures of planet forming disks and potentially even destroy disks completely. \citet{1998ApJ...499..758J} give a disk truncation timescale of $\sim 10^6$ yr within $d \sim 0.3$ pc around O-stars, calibrated using $\theta^1$ Ori C in the Orion Nebula Cluster Trapezium system. Recent models have the capability to accurately calculate disk structures and mass outflows from low to mid background radiation fields \citep{2016MNRAS.457.3593F}, which could be used to elaborate on this issue. With the parameters chosen we are confident that the disks left in our enriched ensemble did not suffer massive outflows by photoevaporation.

This leads to the time dependent evolution of the protoplanetary disks in our models. We chose the classic $\alpha$ viscosity disk model by \citet{2009apsf.book.....H}, which is considered a quasi-standard in the literature and is widely used in observational and theoretical modeling of planet forming disks. Recently, however, doubts about the nature of angular momentum transfer in disks and the physical cause for disk dispersal have arisen \citep{2014prpl.conf..475A}. An additional caveat closely related is that circumstellar disks in our model dissipate on comparable timescales independent of stellar mass. In fact there is clear evidence that circumstellar disks around higher mass stars dissipate more rapidly around stars of higher mass \citep{1998AJ....116.1816H,2006ApJ...651L..49C}.  While one is tempted to assume that characteristic disk lifetimes depend linearly on star mass (qualitatively consistent with available evidence) future work will quantify this relationship and enable a more sophisticated population model than assumed here. 

The disk disruption and disk truncation due to supernova feedback and momentum stripping and ram pressure by supernova ejecta demand more detailed modeling in future work. The values used in this work were derived with the enrichment of the Solar System in mind \citep{2000ApJ...538L.151C,2007ApJ...662.1268O,2010ApJ...711..597O} and are therefore not perfectly transferable to other types of disk structures and stellar parameters. We accounted for that by extrapolating these findings to other disk parameters, however, more detailed modeling is necessary to find robust criteria for disk stripping and mass loss by supernova ejecta.

Finally, the derived heating values reflect the maximum enrichment of a single system, which may be asynchronous to solid condensation and thus incorporation into planetesimals and other planetary precursor material. Therefore, the values in Sect. \ref{sec:enrichment_distribution} can be seen as maximum or `initial' values for enriched systems. However, this did not affect the number of enriched systems in total and demonstrates that enrichment levels can reach extremely high values in comparison with the Solar System. 



\section{Conclusions}
\label{sec:conclusions}

The supernova pollution of forming planetary systems with SLRs has the potential to crucially alter the growth, interior evolution and volatile budget of terrestrial planets. We have conducted numerous \nbody simulations of the evolution of young star clusters of sizes comparable to the Solar birth cluster (10$^3$--10$^4$ stars) with varying morphology and realistic stellar evolution. From these we have derived SLR enrichment levels for circumstellar disks struck by supernova ejecta, which are not affected by shock front disruption, dynamical encounters or intense ambient radiation fields. For these system we calculated initial heating values from radioactive decay under the assumption that planetesimals form soon after the enrichment event. Our conclusions can be summarized as follows:

\begin{itemize}
\item The potential planetary systems exhibited a wide range in their SLR abundances: $\sim$10--30\% of all systems were enriched, bearing high SLR levels, whereas many systems had negligible or zero abundances.
\item Among enriched systems, Solar System SLR enrichment levels were common. However, we do not exactly match the Solar \al/\fe ratio, which is a consquence of our assumption of isotropic and homogenous supernova ejecta.
\item The most extreme heating values could be several orders of magnitude higher than those for the Solar System at CAI formation.
\end{itemize}

We argue that significant SLR levels can have important influence on the early and long-term evolution of planets by altering interior thermo-mechanical evolution and the volatile budget. These mechanisms may crucially determine exoplanet observables, like atmospheric abundances or radius, and habitability and could be reflected in the global galaxy populations.

Future investigations will improve the link between photoevaporative mass loss with stellar winds and more refined disk models. Furthermore, advanced understanding of melt migration and volatile degassing in planetesmals and planetary embryos with varied radiogenic heating rates is needed to quantify consequences for the terrestrial planetary assembly. 

\section*{Acknowledgements}

We thank an anonymous referee for valuable comments that helped to improve the quality of the paper. We gratefuly acknowledge discussions with Gregor J. Golabek, Maria Sch{\"o}nb{\"a}chler, Lee Hartmann and Fred J. Ciesla. TL was supported by ETH Research Grant ETH-17 13-1. RJP acknowledges support from the Royal Astronomical Society in the form of a research fellowship. The numerical simulations in this work were performed on the \textsc{brutus} and \textsc{euler} computing clusters of ETH Z{\"u}rich. The models were analyzed using the open source software environment \textsc{matplotlib}\footnote{\url{http://matplotlib.org}} \citep{matplotlib}. Parts of this work have been carried out within the framework of the National Center for Competence in Research PlanetS supported by the SNSF.





\bibliographystyle{mnras}
\bibliography{bibliography,licht}



\bsp	
\label{lastpage}
\end{document}

%% file: tabs/tab_cluster_results.txt
10$^3$ & Fractal & 1.0 & 2.0 $\pm$ 1.3 & 0.6 $\pm$ 0.4 & 3.5 $\pm$ 1.6 & 17.2 $\pm$ 8.7 & 20.3 $\pm$ 9.6 \\
10$^3$ & Plummer & 0.3 & 1.7 $\pm$ 1.2 & 0.6 $\pm$ 0.5 & 0.1 $\pm$ 0.1 & 26.3 $\pm$ 15.6 & 26.6 $\pm$ 15.7 \\
10$^3$ & Plummer & 0.4 & 1.7 $\pm$ 1.2 & 0.7 $\pm$ 0.7 & 0.1 $\pm$ 0.1 & 25.1 $\pm$ 13.9 & 25.4 $\pm$ 14.1 \\
10$^4$ & Fractal & 1.0 & 17.2 $\pm$ 4.1 & 1.5 $\pm$ 0.6 & 4.3 $\pm$ 0.9 & 49.2 $\pm$ 8.1 & 52.3 $\pm$ 7.5 \\
10$^4$ & Fractal & 3.0 & 18.8 $\pm$ 4.4 & 0.9 $\pm$ 0.3 & 3.2 $\pm$ 0.8 & 30.4 $\pm$ 6.0 & 33.1 $\pm$ 5.7 \\
10$^4$ & Plummer & 0.3 & 20.1 $\pm$ 4.9 & 2.9 $\pm$ 1.0 & 0.1 $\pm$ 0.0 & 78.8 $\pm$ 3.8 & 79.1 $\pm$ 3.7 \\
10$^4$ & Plummer & 0.4 & 20.7 $\pm$ 4.7 & 4.1 $\pm$ 1.4 & 0.0 $\pm$ 0.0 & 68.7 $\pm$ 3.2 & 69.2 $\pm$ 3.2 \\

%% file: tabs/tab_enrichment_stats.txt
10$^3$ & Fractal & 1.0 & 22.9 $\pm$ 10.1 & 3.5 $\pm$ 10.4 & 1.8 $\pm$ 6.7 & 1.2 $\pm$ 1.0 & 6.6 $\pm$ 4.6 \\
10$^3$ & Plummer & 0.3 & 16.8 $\pm$ 9.7 & 6.7 $\pm$ 15.8 & 3.7 $\pm$ 10.5 & 2.3 $\pm$ 1.9 & 9.3 $\pm$ 6.2 \\
10$^3$ & Plummer & 0.4 & 17.4 $\pm$ 10.1 & 6.1 $\pm$ 12.9 & 3.5 $\pm$ 8.2 & 2.2 $\pm$ 1.5 & 10.1 $\pm$ 6.8 \\
10$^4$ & Fractal & 1.0 & 20.3 $\pm$ 3.4 & 9.2 $\pm$ 17.1 & 10.1 $\pm$ 18.7 & 5.7 $\pm$ 2.0 & 12.1 $\pm$ 2.4 \\
10$^4$ & Fractal & 3.0 & 28.3 $\pm$ 2.2 & 8.0 $\pm$ 14.7 & 8.5 $\pm$ 16.1 & 4.8 $\pm$ 1.6 & 16.4 $\pm$ 3.2 \\
10$^4$ & Plummer & 0.3 & 8.9 $\pm$ 1.6 & 15.7 $\pm$ 22.2 & 17.0 $\pm$ 24.2 & 9.3 $\pm$ 2.8 & 7.0 $\pm$ 1.4 \\
10$^4$ & Plummer & 0.4 & 13.3 $\pm$ 1.6 & 23.6 $\pm$ 30.6 & 24.8 $\pm$ 32.0 & 13.9 $\pm$ 3.9 & 11.6 $\pm$ 1.5 \\